\documentclass[aps,prb,reprint,superscriptaddress]{revtex4-2}

\usepackage{graphicx}% Include figure files

\begin{document}

\title{Origin of the 30~T transition in CeRhIn$_5$ in tilted magnetic fields}

\author{S.~Mishra}
\affiliation{Laboratoire National des Champs Magn\'{e}tiques Intenses (LNCMI-EMFL), CNRS, UGA, 38042 Grenoble, France}

\author{D.~Gorbunov}
\affiliation{Hochfeld-Magnetlabor Dresden (HLD-EMFL) and W\"{u}rzburg-Dresden Cluster of Excellence ct.qmat, Helmholtz-Zentrum Dresden-Rossendorf, 01328 Dresden, Germany}

\author{D.J.~Campbell}
\affiliation{Laboratoire National des Champs Magn\'{e}tiques Intenses (LNCMI-EMFL), CNRS, UGA, 38042 Grenoble, France}

\author{D.~LeBoeuf}
\affiliation{Laboratoire National des Champs Magn\'{e}tiques Intenses (LNCMI-EMFL), CNRS, UGA, 38042 Grenoble, France}

\author{J.~Hornung}
\affiliation{Hochfeld-Magnetlabor Dresden (HLD-EMFL) and W\"{u}rzburg-Dresden Cluster of Excellence ct.qmat, Helmholtz-Zentrum Dresden-Rossendorf, 01328 Dresden, Germany}
\affiliation{Institut f\"ur Festk\"orper- und Materialphysik, TU Dresden, 01062 Dresden, Germany}

\author{J.~Klotz}
\affiliation{Hochfeld-Magnetlabor Dresden (HLD-EMFL) and W\"{u}rzburg-Dresden Cluster of Excellence ct.qmat, Helmholtz-Zentrum Dresden-Rossendorf, 01328 Dresden, Germany}

\author{S.~Zherlitsyn}
\affiliation{Hochfeld-Magnetlabor Dresden (HLD-EMFL) and W\"{u}rzburg-Dresden Cluster of Excellence ct.qmat, Helmholtz-Zentrum Dresden-Rossendorf, 01328 Dresden, Germany}

\author{H.~Harima}
\affiliation{Graduate School of Science, Kobe University, Kobe 657-8501, Japan}

\author{J.~Wosnitza}
\affiliation{Hochfeld-Magnetlabor Dresden (HLD-EMFL) and W\"{u}rzburg-Dresden Cluster of Excellence ct.qmat, Helmholtz-Zentrum Dresden-Rossendorf, 01328 Dresden, Germany}
\affiliation{Institut f\"ur Festk\"orper- und Materialphysik, TU Dresden, 01062 Dresden, Germany}

\author{D. Aoki}
\affiliation{Institute for Materials Research, Tohoku University, Oarai, Ibaraki, 311-1313, Japan}

\author{A.~McCollam}
\affiliation{High Field Magnet Laboratory (HFML-EMFL), Radboud University, 6525 ED Nijmegen, The Netherlands}

\author{I.~Sheikin}
\email[]{ilya.sheikin@lncmi.cnrs.fr}
\affiliation{Laboratoire National des Champs Magn\'{e}tiques Intenses (LNCMI-EMFL), CNRS, UGA, 38042 Grenoble, France}

\date{\today}

\begin{abstract}
We present a comprehensive ultrasound study of the prototypical heavy-fermion material CeRhIn$_5$, examining the origin of the enigmatic 30~T transition. For a field applied at 2$^\circ$ from the $c$ axis, we observed two sharp anomalies in the sound velocity, at $B_m \approx$~20~T and $B^* \approx$~30~T, in all the symmetry-breaking ultrasound modes at low temperatures. The lower-field anomaly corresponds to the well-known first-order metamagnetic incommensurate-to-commensurate transition. The higher-field anomaly takes place at 30~T, where an electronic-nematic transition was previously suggested to occur. Both anomalies, observed only within the antiferromagnetic state, are of similar shape, but the corresponding changes of the ultrasound velocity have opposite signs. Based on our experimental results, we suggest that a field-induced magnetic transition from a commensurate to another incommensurate antiferromagnetic state occurs at $B^*$. With further increasing the field angle from the $c$ axis, the anomaly at $B^*$ slowly shifts to higher fields, broadens, and becomes smaller in magnitude. Traced up to 30$^\circ$ from the $c$ axis, it is no longer observed at 40$^\circ$ below 36~T.
\end{abstract}

\maketitle

\section{Introduction}

Unusual magnetic-field-induced transitions have recently become a matter of considerable interest in strongly correlated electron systems. In some heavy-fermion compounds, intermetallic materials based on rare earths or actinides, such transitions are rather exotic in nature. Among them are field-induced quantum critical points~\cite{Gegenwart2002,Harrison2007}, field-induced or reinforced unconventional superconductivity~\cite{Levy2005,Aoki2009,Knebel2019}, and Lifshitz transitions~\cite{Daou2006,Rourke2008,Aoki2016}. Understanding the mechanisms and the origins of such transitions is of primary importance in modern condensed matter.

\begin{figure}[ht]
  \includegraphics[width=\columnwidth]{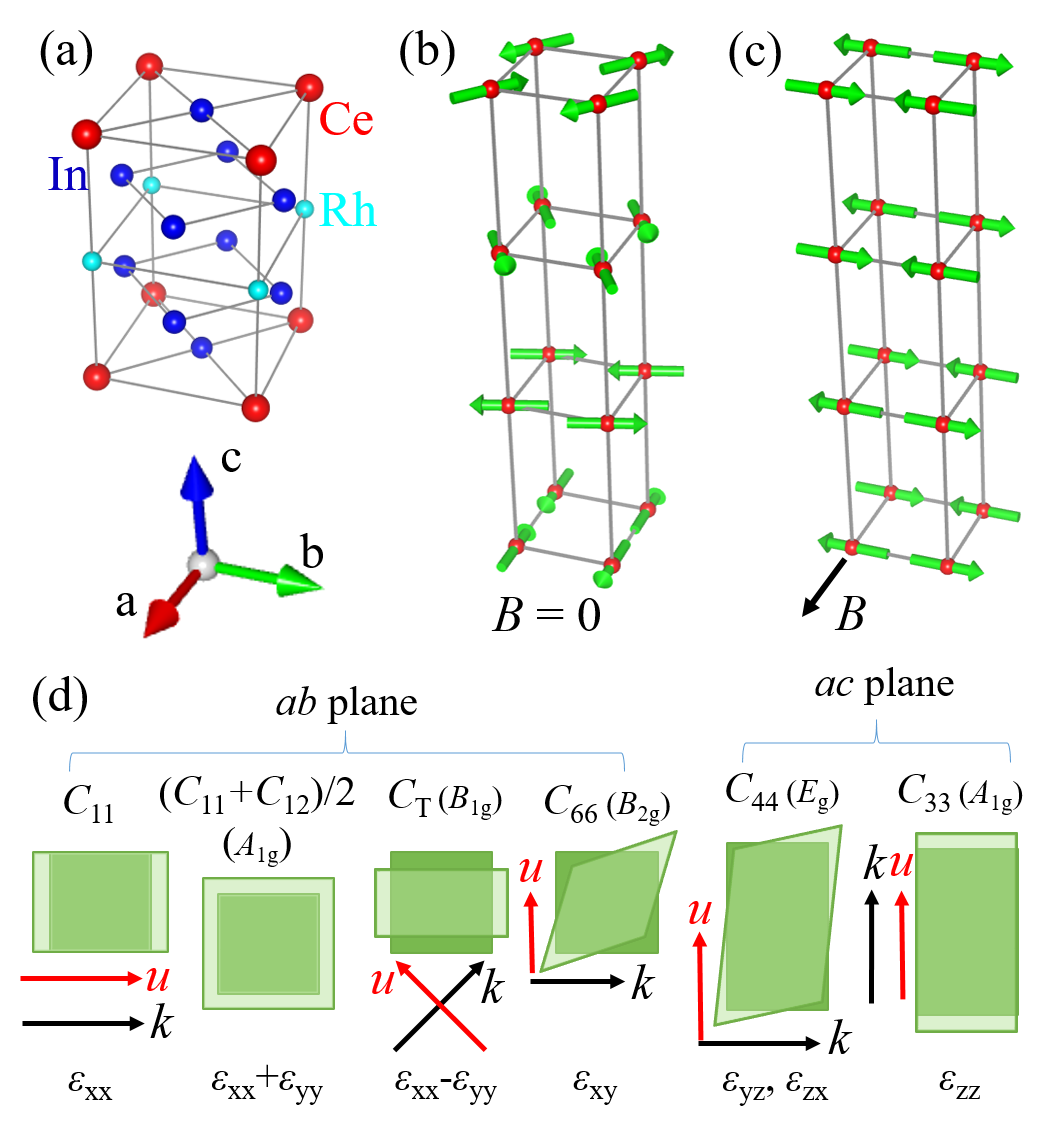}
  \caption{(a) Crystal structure of CeRhIn$_5$. Magnetic structure of CeRhIn$_5$ (b) in zero magnetic field (IC) and (c) in a magnetic field higher than 2~T applied along the [100] direction (C). Arrows indicate the orientation of the magnetic moments. Only Ce atoms are shown in (b) and (c) for clarity. (d) Schematic illustration of the symmetry strains, $\varepsilon_{ij}$, induced by different ultrasound modes for a tetragonal crystal structure. For each mode, the propagation ($k$) and polarization ($u$) directions are shown by arrows. The associated irreducible representations are shown in brackets.}
  \label{fig:Structure}
\end{figure}

CeRhIn$_5$, although discovered barely two decades ago~\cite{Hegger2000}, is one of the best studied heavy-fermion materials. This compound crystallizes into a tetragonal HoCoGa$_5$ type crystal structure (space group 123, $P4/mmm$) shown in Fig.~\ref{fig:Structure}(a). CeRhIn$_5$ undergoes an antiferromagnetic (AFM) transition at $T_N =$~3.8~K. The zero-field magnetic structure, AFM1, shown in Fig.~\ref{fig:Structure}(b), is incommensurate (IC) with propagation vector $\mathbf{Q}_{IC} = (1/2, 1/2, 0.297)$~\cite{Bao2000,Fobes2017}. The Ce magnetic moments are antiferromagnetically aligned within the basal plane, and spiral transversally along the $c$ axis.

At low temperatures, when a magnetic field $B$ is applied in the basal plane, a first-order transition occurs at $B_m \simeq$~2~T. The transition corresponds to a change of magnetic structure from IC AFM1 to commensurate (C) AFM3 with the propagation vector $\mathbf{Q}_{C} = (1/2, 1/2, 1/4)$~\cite{Raymond2007,Fobes2018}. It is a collinear square configuration with ``up-up-down-down'' alignment. The magnetic moments are antiferromagnetically aligned in the basal plane along the direction perpendicular to the magnetic field, as shown in Fig.~\ref{fig:Structure}(c). When the magnetic field is tilted away from the basal plane, $B_m$ shifts to higher fields and follows the $1/\cos\alpha$ dependence ($\alpha$ is the field angle from the basal plane) up to about 75$^\circ$--80$^\circ$. At higher angles, however, $B_m$ deviates from a $1/\cos\alpha$ dependence towards lower fields. The transition was traced up to $\theta = 2^\circ$ from the $c$ axis ($\alpha =$~88$^\circ$), where it occurs at $B_m \approx$~20~T~\cite{Mishra2021}.

Recent results obtained in high magnetic fields applied along or close to the $c$ axis suggest a remarkably novel behavior in CeRhIn$_5$. The AFM order is suppressed at $B_c \simeq$~50~T giving rise to a field-induced quantum critical point~\cite{Jiao2015,Jiao2019}. Furthermore, Moll \textit{et al.}~\cite{Moll2015} reported an observation of a hysteretic jump in the in-plane resistivity of CeRhIn$_5$ microstructures fabricated by focused ion beam (FIB) at $B^{*}\approx$ 30~T applied along or close to the $c$ axis. Later on, these experiments were advanced, allowing for simultaneous measurements of the in-plane resistivity along the [100] and [010], as well as along the [110] and [1$\overline{1}$0] directions~\cite{Ronning2017}. These measurements revealed a strong in-plane resistivity anisotropy within each of the two inequivalent symmetry channels $B_{1g}$ ([100], [010]) and $B_{2g}$ ([110], [1$\overline{1}$0]), which emerges above $B^{*}\approx$ 30~T tilted from the $c$ axis towards the [100] and [110] directions, respectively. This electronic anisotropy was interpreted in terms of an electronic-nematic transition. To the best of our knowledge, these results, however, have not been reproduced on bulk samples so far.

For a long time, all the attempts to detect an anomaly at $B^*$ in various measurements on bulk single crystals remained ineffective. In particular, no noticeable anomaly was observed at $B^*$ in longitudinal magnetization~\cite{Takeuchi2001} or reported in magnetic-torque data~\cite{Moll2015,Ronning2017}. It is only very recently that a small magnetostriction anomaly was observed at $B^*$ in a magnetic field applied at ${\sim}20^\circ$ from the $c$ axis~\cite{Rosa2019}. The anomaly is similar in shape and size to the one at $B_m$, which occurs at about 7.5~T for this field orientation. Furthermore, recent high-field nuclear magnetic resonance (NMR) measurements revealed a pronounced decrease in the $^{115}$In Knight shift at $B^*$ for $B \parallel c$~\cite{Lesseux2020}. These NMR measurements also suggest an IC magnetic structure of CeRhIn$_5$ above $B^*$. Both the magnetostriction anomaly and the change of the NMR Knight shift at $B^*$ were interpreted as a signature of enhanced hybridization between the 4$f$ and conduction electrons. This interpretation is in line with the conclusion of an earlier work, which suggests that the Ce $f$ electrons, localized at low fields, become itinerant above $B^*$~\cite{Jiao2015,Jiao2017}. This scenario, however, was later ruled out by high-field angular-dependent de Haas-van Alphen (dHvA) effect measurements, which demonstrated that the $f$ electrons remained localized not only above $B^*$ but also above $B_c$~\cite{Mishra2021}. Finally, a specific-heat anomaly has recently been observed at $B^*$ for $B \parallel c$, suggesting that a true thermodynamic phase transition, probably weakly first order, takes place at $B^*$~\cite{Mishra2021a}.

The exact origin of the transition at $B^*$, especially in bulk crystals, remains, therefore, an important open question. One way to address this issue is via ultrasound-velocity measurements, which are a powerful tool to detect phase transitions since ultrasound waves readily couple with electronic, magnetic, and structural degrees of freedom. Moreover, different ultrasound modes, characterized by different propagation directions and polarization, induce different symmetry-breaking strains in the crystal, as schematically shown in Fig.~\ref{fig:Structure}(d) for a tetragonal crystal structure. This, in turn, allows for the determination of the order-parameter symmetry in a phase as well as identification of broken symmetries at a phase transition.

Previous ultrasound velocity measurements performed in pulsed magnetic fields slightly tilted from the $c$ axis revealed clear anomalies at both $B_m \approx$~20~T and $B^* \approx$~30~T~\cite{Kurihara2020}. The latter was observed in the $C_T = (C_{11} - C_{12})/2$ mode only, which induces the symmetry strain $\varepsilon_{xx}-\varepsilon_{yy}$ associated with the irreducible representation $B_{1g}$. This strain breaks the $C_4$ symmetry [Fig.~\ref{fig:Structure}(d)]. Another strain, which breaks the $C_4$ symmetry, is $\varepsilon_{xy}$. This strain, induced in the $C_{66}$ mode, is associated with the $B_{2g}$ symmetry channel [Fig.~\ref{fig:Structure}(d)]. If the anomaly at $B^*$ is attributed to the previously suggested electronic-nematic transition~\cite{Ronning2017}, it is surprising that no anomaly was observed in the $C_{66}$ and $C_{11}$ modes. These inconsistencies motivated us to re-examine the elastic response of CeRhIn$_5$ in high magnetic fields.

In this paper, we report high-resolution ultrasound-velocity measurements in CeRhIn$_5$ in high pulsed and static magnetic fields tilted away from the $c$ axis. We observed clear field-induced anomalies at both $B_m$ and $B^*$ in all symmetry-breaking modes. This suggests that the transition at $B^*$ is bulk in origin contrary to what was previously suggested. Both anomalies are of similar shape, but of the opposite sign. Both of them exist within the AFM state only. Furthermore, when the magnetic field is tilted further away from the $c$ axis, the anomaly at $B^*$ becomes broader and smaller in magnitude. It was traced up to an angle tilted by 30$^\circ$ away from the $c$ axis.

\section{Experimental details}

High-quality single crystals of CeRhIn$_5$ used in this study were grown by the In-self-flux technique, details of which can be found elsewhere~\cite{Shishido2002}. Ultrasound velocity measurements were performed in both pulsed (up to 60~T) and static (up to 36~T) magnetic fields. In pulsed fields, we used a $^4$He flow cryostat, while static-field measurements were performed in a $^3$He cryostat equipped with an \emph{insitu} rotator. In both cases, the measurements were performed using a pulse-echo technique in transmission mode. A pair of piezoelectric transducers was glued to opposite well-polished surfaces of the sample in order to generate and detect acoustic waves. The ultrasound frequencies and the absolute values of sound velocities at 4.2~K for different acoustic modes are given in Table~\ref{tab:parameters}.

\begin{table}[hbt]%\centering
\caption{\label{tab:parameters}Ultrasound frequencies and velocities at $T =$~4.2~K for all measured acoustic modes. The error in ultrasound velocities is estimated to be 80 m/s.}
\begin{ruledtabular}
\begin{tabular}{c c c c c c}

   & $C_{11}$ & $C_T$ & $C_{33}$ & $C_{44}$ & $C_{66}$ \\
   \hline
  $f$ (MHz) & 115.6 & 134 & 104 & 173 & 105 \\
  $v$ (m/s) & 3514 & 2127 & 4484  & 2083 & 2016 \\

\end{tabular}
\end{ruledtabular}
\end{table}

For all modes except $C_{33}$, the orientation of the magnetic field with respect to the $c$ axis was determined from the position of the anomaly at $B_m$, whose angular dependence was established in the previous torque measurements~\cite{Mishra2021}. For the $C_{33}$ mode, the orientation was determined from the frequencies of the magnetoacoustic quantum oscillations, which were compared to the previously established angular dependence of the de Haas-van Alphen (dHvA) frequencies~\cite{Mishra2021}. The oscillatory frequencies also allowed us to double check the field angle in other modes, in which the magnetoacoustic oscillations were strong enough.

\section{Results and discussion}

\begin{figure}[ht]
  \includegraphics[width=\columnwidth]{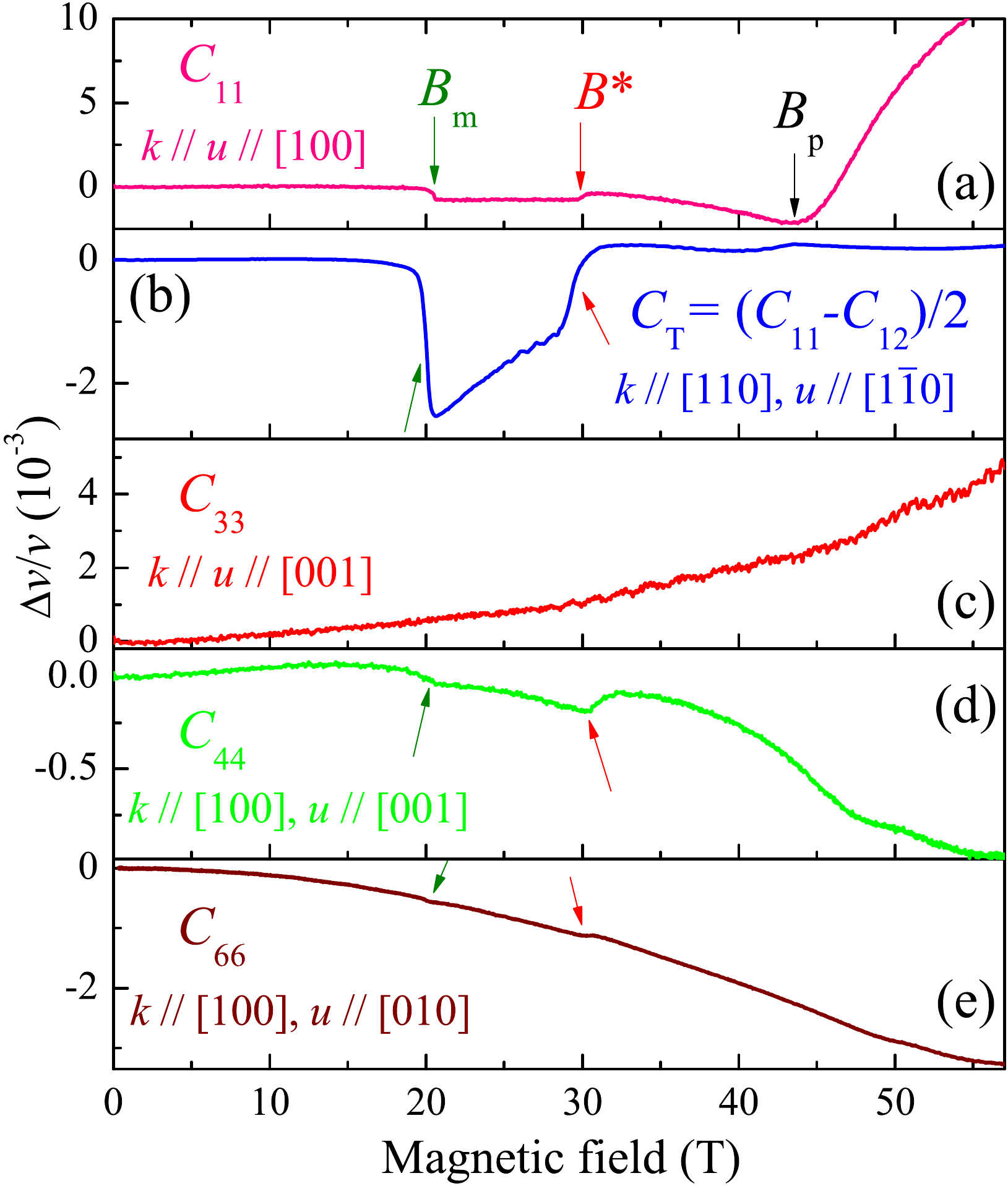}
  \caption{Field dependence of the relative ultrasound-velocity variation, $\Delta v/v$, for the $C_{11}$ (a), $C_T = (C_{11} - C_{12})/2$ (b), $C_{33}$ (c), $C_{44}$ (d), and $C_{66}$ (e) modes at $T \simeq$~1.4~K.  The ultrasound frequencies for all modes are given in Table~\ref{tab:parameters}. The field was applied at 2$^\circ$ from the $c$ axis for all  modes, except for the $C_{33}$ mode, where the angle was 4$^\circ$. Arrows indicate the anomalies at $B_m$, $B^*$, and $B_p$ discussed in the text. For all the modes, the ultrasound propagation $k$ and polarization $u$ directions are indicated.}\label{fig:All_modes}
\end{figure}

Figure~\ref{fig:All_modes} shows low-temperature relative ultrasound velocity, $\Delta v/v$, as a function of $B$ applied close to the $c$ axis for several ultrasound modes. Two distinct anomalies are observed at $B_m \approx$~20~T and $B^* \approx$~30~T in all symmetry-breaking modes, i.e., $C_{11}$, $C_{44}$, $C_{66}$, and $C_T$. Both anomalies are absent in the $C_{33}$ mode, which does not break any tetragonal symmetry.

\begin{figure}[htb]
\includegraphics[width=\columnwidth]{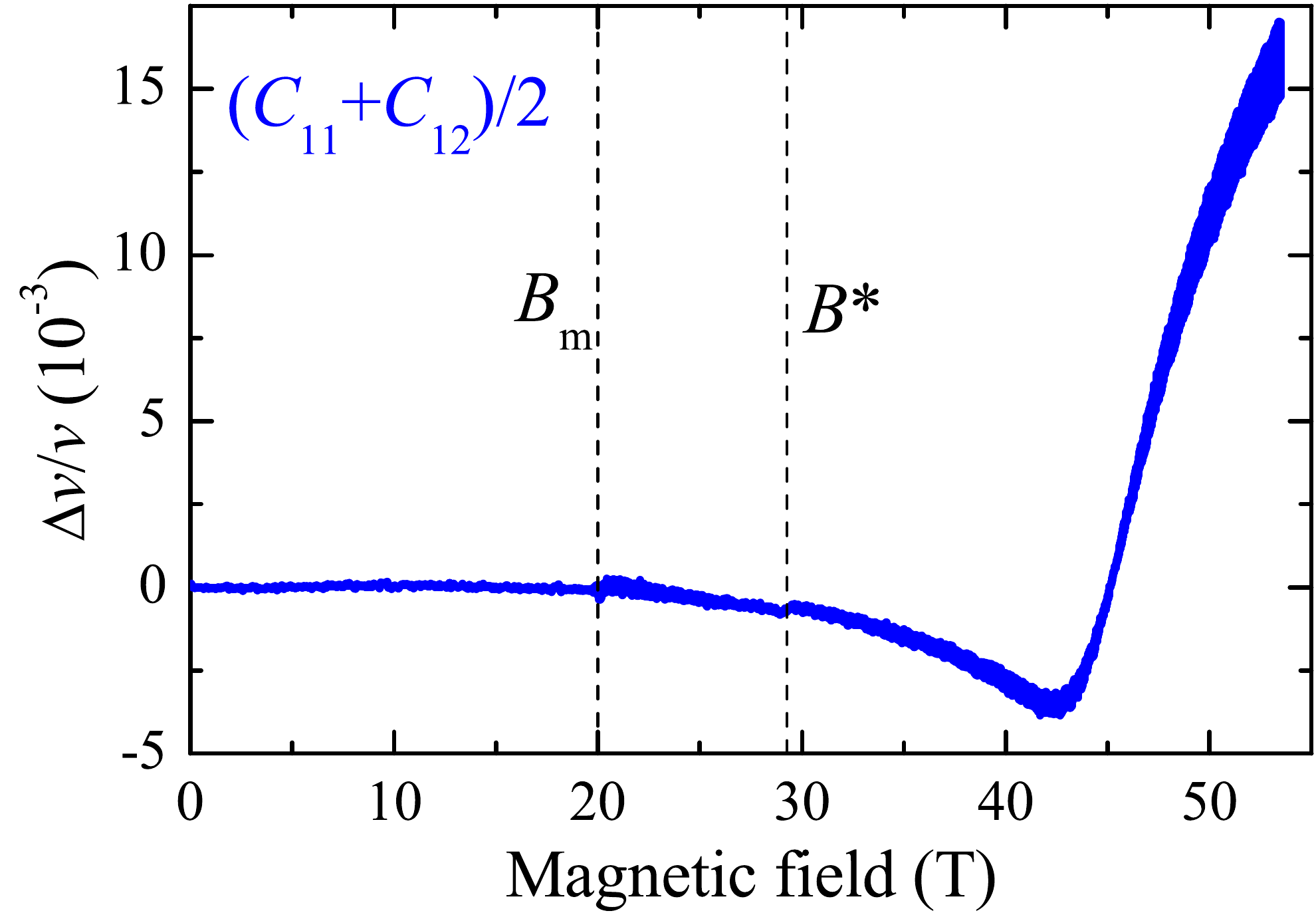}
\caption{\label{fig:New_mode}Field dependence of the relative ultrasound-velocity variation, $\Delta v/v$, for the $(C_{11} + C_{12})/2$ mode at $T \simeq$~1.4~K, calculated from the $C_{11}$ and $C_T$ data shown in Figs.~\ref{fig:All_modes}(a) and \ref{fig:All_modes}(b), respectively, as explained in the text. The field was applied at 2$^\circ$ from the $c$ axis. Dashed lines indicate $B_m$ and $B^*$ observed in the symmetry-breaking modes.}
\end{figure}

The non-symmetry-breaking $(C_{11}+C_{12})/2$ mode related to the $\varepsilon_{xx}+\varepsilon_{yy}$ strain belonging to the $A_{1g}$  representation can not be measured directly using the pulse-echo technique. The relative ultrasound velocity variation for this mode was, therefore, obtained from the $C_{11}$ and $C_T$ data. First, the absolute values of the zero-field elastic constants were calculated as $C_0 = \rho v^2$ using the ultrasonic velocities $v$ given in Table~\ref{tab:parameters} and the density $\rho =$ 8.316 g/cm$^3$. Then, the field-dependent elastic constant variations were calculated as $\Delta C(B) = 2 \frac{\Delta v}{v}(B) C_0$ using the measured field dependence of $\Delta v/v$ for both $C_{11}$ and $C_T$ modes, shown in Figs.~\ref{fig:All_modes}(a) and \ref{fig:All_modes}(b), respectively. Since $(C_{11} + C_{12})/2 = C_{11} - C_T$ and $\Delta (C_{11} + C_{12})/2 = \Delta C_{11} - \Delta C_T$, the field dependence of $\Delta C/C$ for the $(C_{11}+C_{12})/2$ mode was obtained. Finally, the relative ultrasound velocity variation was calculated as $\Delta v/v = (\Delta C/C)/2$.

The resulting velocity variation for the $(C_{11}+C_{12})/2$ mode is shown in Fig.~\ref{fig:New_mode}. Here, the line thickness represents the resulting error due to the uncertainties in the determination of the ultrasound velocities for the $C_{11}$ and $C_T$ modes. One can see that, similar to the $C_{33}$ mode, the $(C_{11}+C_{12})/2$ mode does not show any apparent anomalies within the experimental error.

As mentioned above, both anomalies in the symmetry-breaking modes have a similar shape, but the corresponding changes of $\Delta v/v$ are of opposite sign. In the $C_{11}$ and $C_T$ modes, the anomalies at $B_m$ and $B^*$ appear as a sharp decrease and increase of the ultrasound velocity, respectively. These anomalies are remarkably similar to those previously observed in other compounds, in which field-induced spin-reorientation transitions occur~\cite{Luethi2007,Danshin1987,Fil1991,Zherlitsyn1993,Zherlitsyn1995}. In the $C_{44}$ and $C_{66}$ modes, both characteristic fields result in similar, but somewhat reduced features. Another anomaly, a minimum in $\Delta v/v$ versus $B$, is clearly visible at $B_p \approx$~44~T in the $C_{11}$ mode. This anomaly was also observed in the previous ultrasound study, where it was interpreted as a transition (or crossover) into the polarized paramagnetic state~\cite{Kurihara2020}.

The anomaly at $B_m$ corresponds to the well-known first-order metamagnetic transition, where the magnetic structure changes from the IC AFM1 to the C AFM3 phase, as schematically shown in Fig.~\ref{fig:Structure}. Since the transition is of the first order, it is not surprising that it manifests itself in most of the ultrasound modes.

\begin{figure}[h!]
\includegraphics[width=\columnwidth]{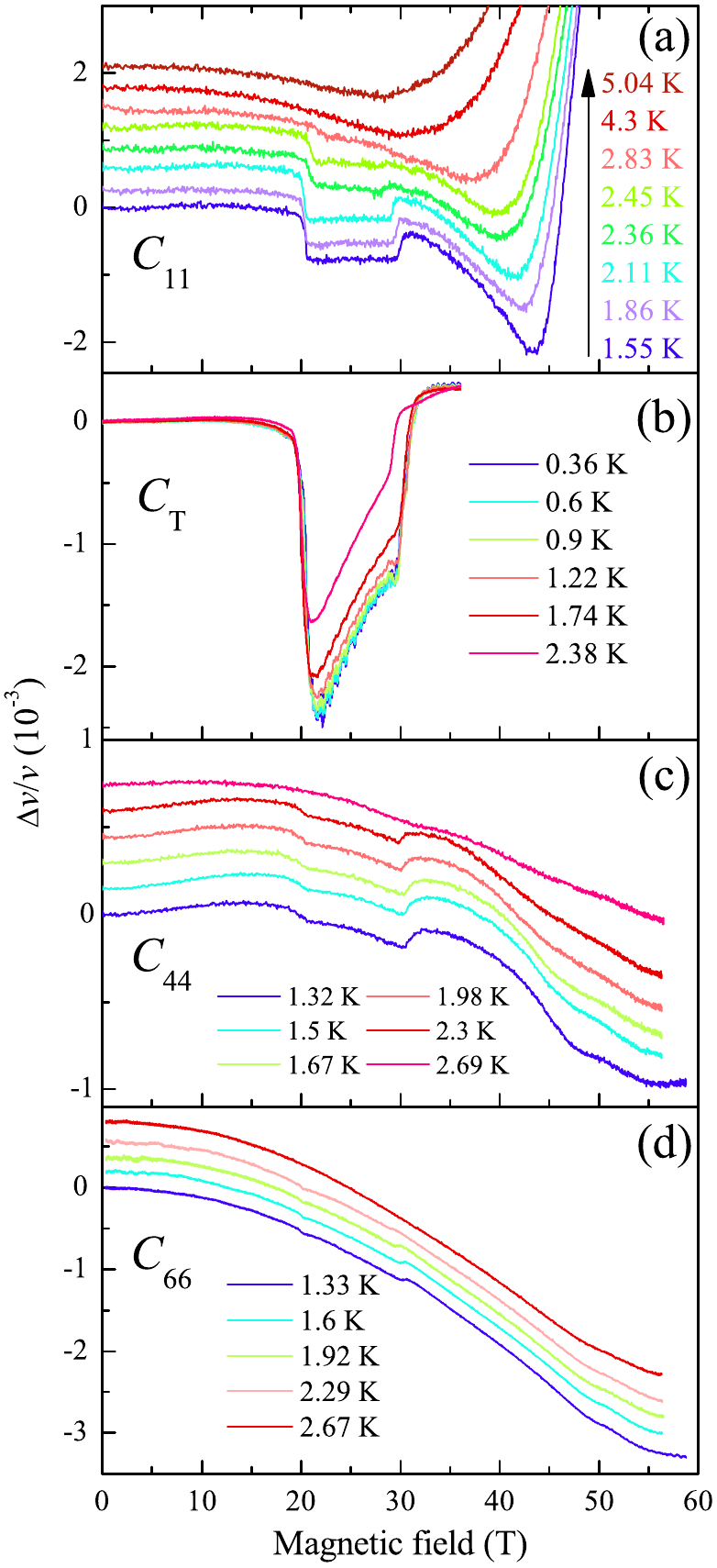}
\caption{\label{fig:T-dependence}Relative variation of the ultrasound velocity as a function of magnetic field at different temperatures for the (a) $C_{11}$, (b) $C_T = (C_{11} - C_{12})/2$, (c) $C_{44}$, and (c) $C_{66}$ acoustic modes. The magnetic field was applied at $\theta \simeq2^\circ$. Curves are vertically shifted in (a), (c), and (d) for clarity.}
\end{figure}

Regarding the transition at $B^*$, its observation in all symmetry-breaking modes suggests that the transition is probably also of the first order, in agreement with recent specific heat measurements~\cite{Mishra2021a}. As mentioned, the anomalies at $B_m$ and $B^*$ have similar shapes but opposite signs. It is, therefore, natural to conclude that both transitions are of the same origin, i.e., that the transition at $B^*$ corresponds to another field-induced change of the magnetic structure, this time from the C AFM3 to an IC AFM4 phase. This conclusion is in line with previous NMR results, which unambiguously suggest an IC phase above $B^*$~\cite{Lesseux2020}.

The above hypothesis regarding the origin of the transition at $B^*$ is further supported by the measurements performed at different temperatures, as shown in Fig.~\ref{fig:T-dependence} for the $C_{11}$ (a), $C_T = (C_{11} - C_{12})/2$ (b), $C_{44}$ (c), and $C_{66}$ (d) acoustic modes. In all cases, the magnetic field was applied at $\theta \simeq 2^\circ$. The data for the $C_T$ mode [Fig.~\ref{fig:T-dependence}(b)] were obtained in static fields. All the other curves were measured in pulsed magnetic fields.

The anomalies at both $B_m$ and $B^*$ remain sharp up to the highest temperature, at which they were observed. The anomaly at $B_m$ is observed up to higher temperatures than its counterpart at $B^*$. Both anomalies are almost temperature independent at low temperatures. At temperatures approaching the antiferromagnetic phase boundary ($\sim$2.5~K for $B_m$ and $\sim$2.4~K for $B^*$), $B_m$ and $B^*$ shift to slightly higher and lower field, respectively.

\begin{figure}[htb]
  \includegraphics[width=\columnwidth]{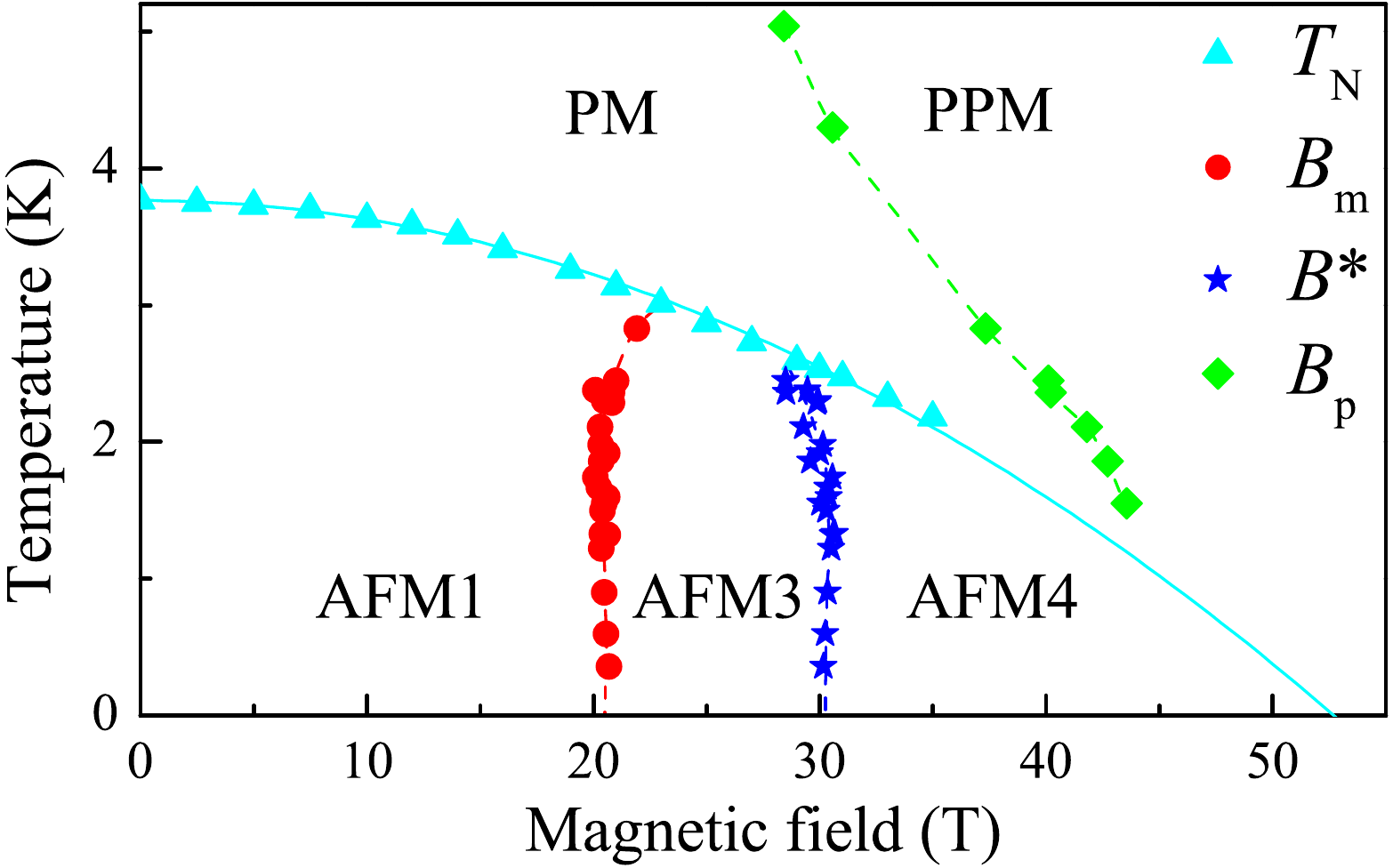}
  \caption{$B\--T$ phase diagram of CeRhIn$_5$ obtained from the ultrasound-velocity anomalies observed in all modes. Triangles correspond to the AFM to paramagnetic (PM) transition from Ref.~\cite{Mishra2021a}. AFM1 and, presumably, AFM4 are IC phases with different propagation vectors. AFM3 is the C phase discussed in the text. PPM stands for the polarized paramagnetic phase. The solid line is a fit of $T_N(B) = T_N(0)[1 - (B/B_c)^2]$ to the N\'{e}el temperature, with $B_c \simeq$~52~T. Dashed lines are guides for the eye.}
  \label{fig:PhaseDiagram}
\end{figure}

\begin{figure}[htb]
  \includegraphics[width=\columnwidth]{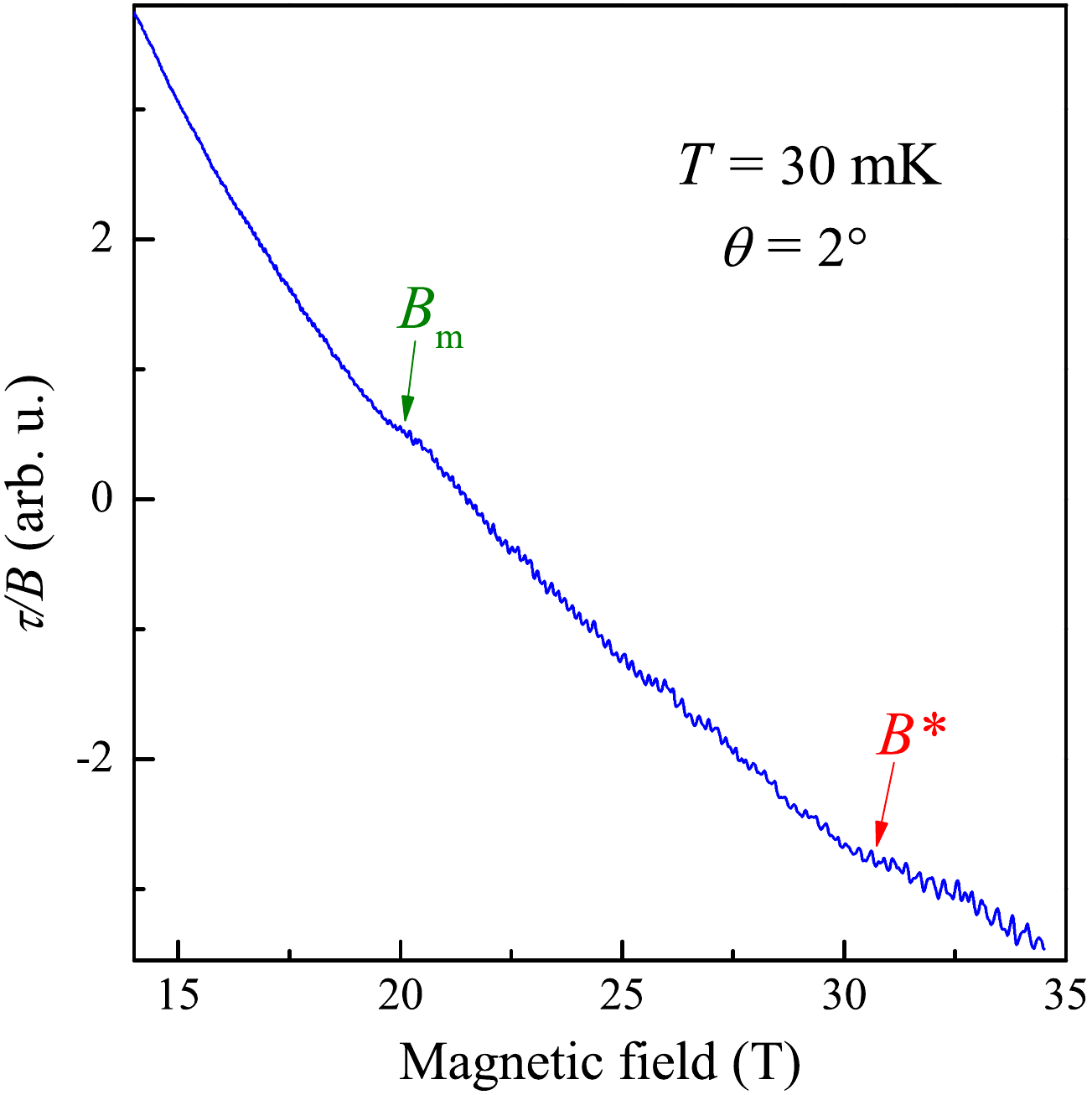}
  \caption{Magnetic torque divided by field, $\tau/B$, vs $B$ applied at $\theta = 2^\circ$ from [001] towards the [110] direction at $T =$~30~mK.}
  \label{fig:Torque}
\end{figure}

Both anomalies from all the measurements are traced in the resulting phase diagram together with the AFM boundary from the previous specific-heat results~\cite{Mishra2021a} (see Fig.~\ref{fig:PhaseDiagram}). In agreement with previous reports~\cite{Ronning2017,Rosa2019,Kurihara2020}, both anomalies at $B_m$ and $B^*$ are observed within the AFM state only, suggesting that both transitions are related to the magnetic properties of CeRhIn$_5$. This assumption is strongly supported by the presence of clear anomalies, both at $B_m$ and $B^*$, in magnetic torque divided by field, $\tau/B$, shown in Fig.~\ref{fig:Torque}.

\begin{figure}[ht]
  \includegraphics[width=\columnwidth]{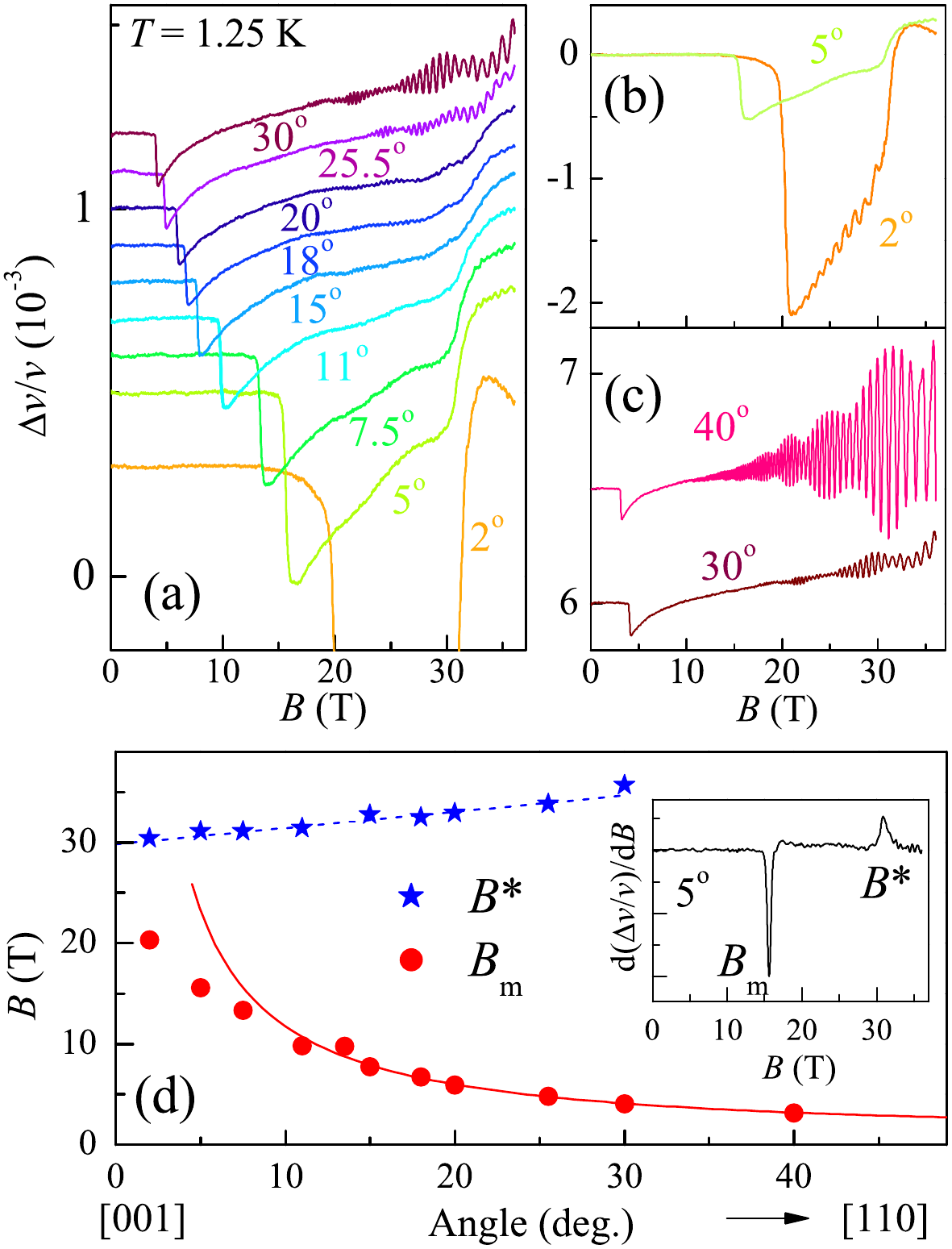}
  \caption{(a) $\Delta v/v$ for the $C_T$ mode as a function of $B$ applied at different angles, $\theta$, from the $c$ axis at $T =$~1.25~K. Curves are vertically shifted for clarity. (b) and (c) show the curves at the two smallest and largest angles of our measurements, respectively. (d) The angle dependence of the transition fields $B_m$ and $B^*$ determined from the minima (maxima) of the first derivative, $d(\Delta v/v)/dB$, an example of which is shown in the inset. The solid line is a $1/\cos(90^\circ - \theta)$ fit to the data above 10$^\circ$. The dashed line is a guide for the eye.}
  \label{fig:Ang-dependence}
\end{figure}

Next, we discuss the angular dependence of the transition fields $B_m$ and $B^*$. The latter was obtained from measurements of the $C_T$ mode, where both anomalies are most pronounced. The field dependence of $\Delta v/v$ for different angles $\theta$ from [001] towards the [110] direction is shown in Fig.~\ref{fig:Ang-dependence}(a). At 2$^\circ$, both anomalies manifest themselves as sharp steps~ [Fig.~\ref{fig:Ang-dependence}(b)]. With increasing angle, both features progressively become smaller. While the anomaly at $B_m$ remains sharp, the one at $B^*$ broadens with increasing angle, {as evidenced by a progressive small increase of the FWHM of the first derivative shown in the inset of Fig.~\ref{fig:Ang-dependence}(d). This probably implies that the transition at $B^*$ at small angles changes to a crossover at larger angles. This assumption is supported by recent specific-heat~\cite{Mishra2021a} and magnetostriction~\cite{Rosa2019} measurements. While the specific heat, performed in magnetic fields applied parallel or very close to the $c$ axis, suggests a phase transition, the magnetostriction, performed in fields tilted by $\sim$20$^\circ$, suggests a crossover at $B^*$. The anomaly at $B_m$ is easily traceable all the way up to 40$^\circ$, while its counterpart at $B^*$ is still present at 30$^\circ$, but is no longer visible at 40$^\circ$, where the curve is completely dominated by strong low-frequency quantum oscillations [Fig.~\ref{fig:Ang-dependence}(c)]. The angular dependence of the anomaly at $B^*$ is strikingly different from that observed in resistivity measurements on FIB-fabricated microdevices~\cite{Moll2015,Ronning2017}, where the resistivity jump at $B^*$ remained sharp all the way up to 60$^\circ$, the highest angle to which it was traced. The size of the jump showed a nonmonotonic behavior with a maximum at about 20$^\circ$ from the $c$ axis. This difference between bulk samples and microfabricated devices is probably due to uniaxial stresses or strains inevitably present in FIB-fabricated devices~\cite{Bachmann2019}. CeRhIn$_5$ seems to be very sensitive to uniaxial strains, especially  to orthorhombic strains, as evidenced by a particularly large size of the anomalies in the $C_T$ mode. Furthermore, previously reported NQR measurements on bulk and powder samples of CeRhIn$_5$ suggest that even small strains (or stresses) change the zero-field magnetic structure from IC to C~\cite{Yashima2020}. Additional measurements under uniaxial stress (or strain) at high magnetic fields on bulk samples are required to elucidate the role of the uniaxial stress in previous transport measurements on CeRhIn$_5$ microstructures~\cite{Moll2015,Ronning2017}.

The resulting angular dependence of both $B_m$ and $B^*$ is shown in Fig.~\ref{fig:Ang-dependence}(d). In agreement with previous results~\cite{Mishra2021}, $B_m$ is strongly angle dependent. It is well fit by $1/\cos(90^\circ - \theta)$ down to about 10$^\circ$ from the $c$ axis but deviates at lower angles. $B^*$, on the other hand, is only weakly angle dependent. It increases approximately linearly all the way from 2$^\circ$ to about 30$^\circ$, above which it either disappears or shifts to beyond 36~T. The remarkable difference between the two angular dependencies is probably due to different magnetic moment arrangements in the vicinity of the two transitions. While the moments are aligned in the basal plane in the vicinity of $B_m$, they are probably tilted towards the $c$ axis close to $B^*$.

Finally, we comment on the magnetoacoustic quantum oscillations. In agreement with the results of the previous high-field ultrasound study of CeRhIn$_5$~\cite{Kurihara2020}, we observed strong low-frequency magnetoacoustic quantum oscillations in the $C_T$ mode between $B_m$ and $B^*$, as shown in the inset of Fig.~\ref{fig:MAQOCTstatic1p25K}. In this mode, these oscillations with a frequency denoted $A$ ($F_A \simeq$~0.6~kT) disappear above $B^*$, as shown in Fig.~\ref{fig:MAQOCTstatic1p25K}, also in agreement with the previous study. This experimental result was interpreted as evidence of a possible Fermi-surface reconstruction at $B^*$ in the previous work~\cite{Kurihara2020}. Similarly, additional oscillatory frequencies emerging above $B^*$ were observed in the dHvA effect~\cite{Jiao2015,Jiao2017} and magnetostriction~\cite{Rosa2019}.  The appearance of these additional frequencies was discussed in terms of the Fermi-surface reconstruction at $B^*$ due to $f$-electrons delocalization~\cite{Jiao2015,Jiao2017}.

In Ref.~\cite{Kurihara2020}, quantum oscillations were observed in the $C_T$ mode only, and only between $B_m$ and $B^*$. In our experiments, on the contrary, we observed magnetoacoustic quantum oscillations in all modes both in static and pulsed-field measurements. Furthermore, the oscillations are the strongest in the $C_{33}$ mode.

\begin{figure}[htb]
\includegraphics[width=\columnwidth]{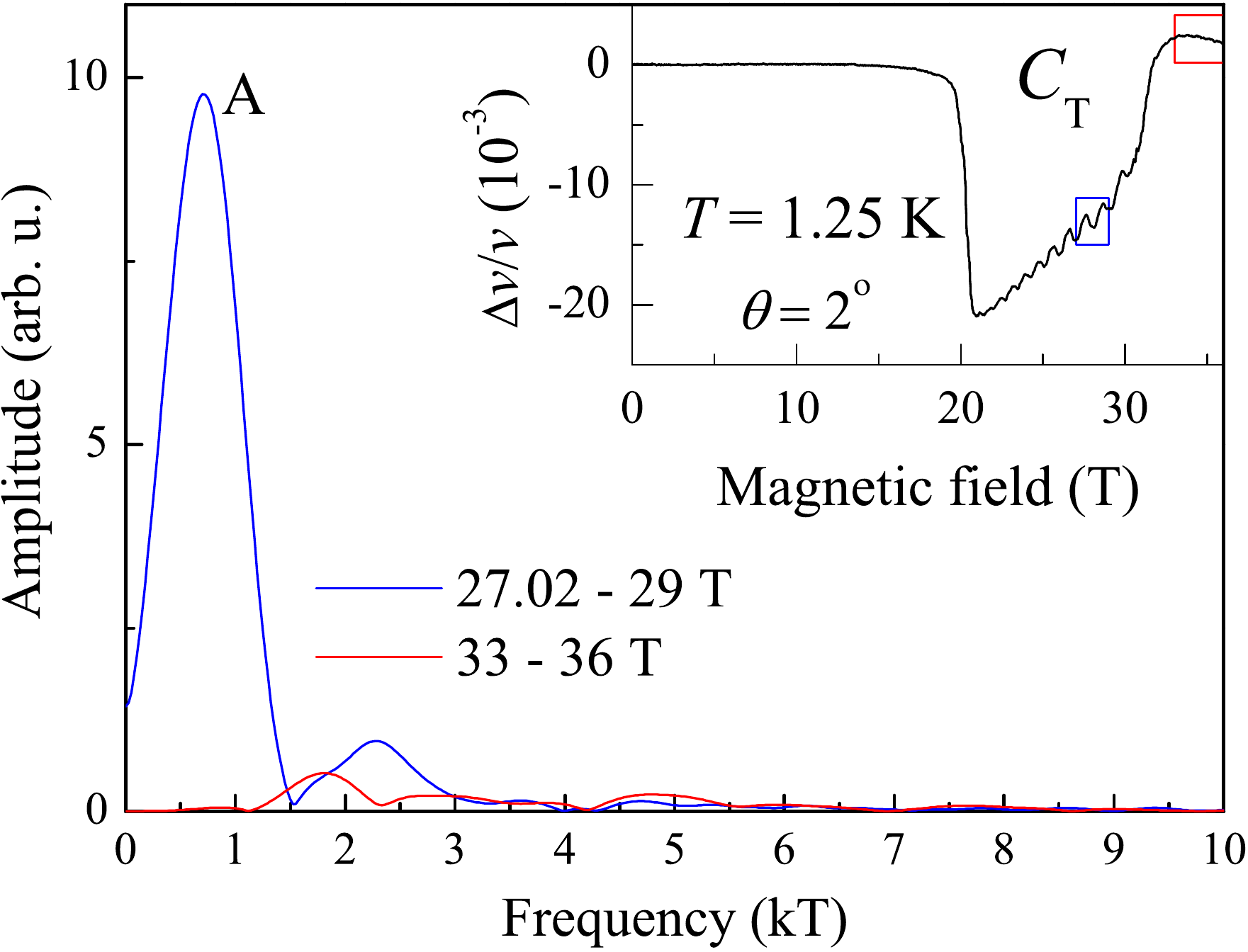}
\caption{\label{fig:MAQOCTstatic1p25K}Fast Fourier transform (FFT) spectra of the magnetoacoustic quantum oscillations in the $C_T$ mode (shown in the inset) both below and above $B^*$. A nonoscillating background was subtracted prior to performing the FFTs. The FFTs were performed over equidistant $1/B$ ranges indicated by rectangles in the inset.}
\end{figure}

\begin{figure}[h!]
\includegraphics[width=\columnwidth]{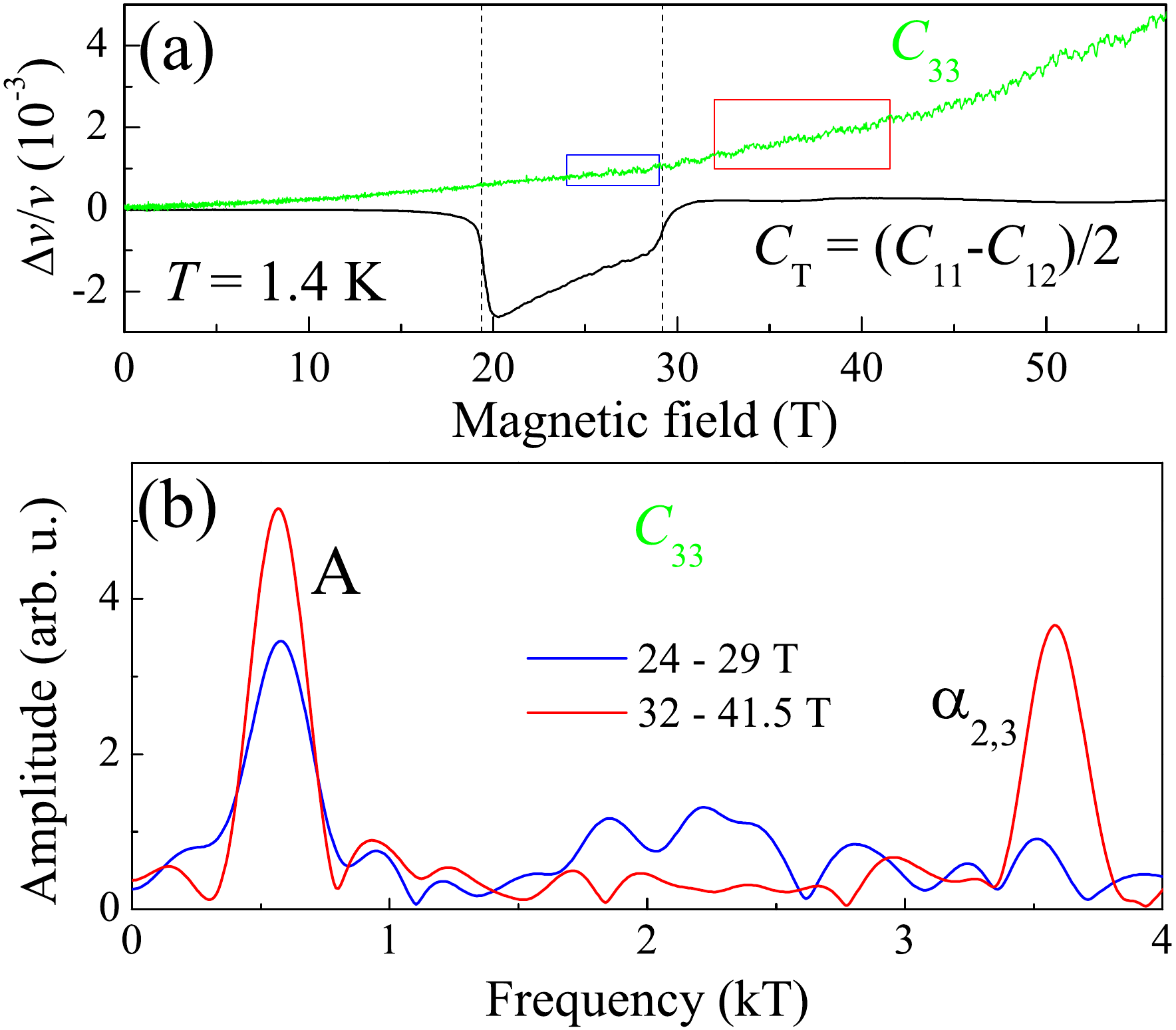}
\caption{\label{fig:MAQOC33pulsed}(a) Field dependence of the relative ultrasound-velocity variation for the modes $C_{33}$ and $C_T$ with the field applied at $\theta = 4^\circ$ and 2$^\circ$, respectively. (b) FFT spectra of the magnetoacoustic quantum oscillations in the $C_{33}$ mode from (a), both below and above $B^*$. A nonoscillating background was subtracted prior to performing the FFTs. The FFTs were performed over equidistant $1/B$ ranges indicated by rectangles in (a).}
\end{figure}

Figure~\ref{fig:MAQOC33pulsed}(a) shows the field dependence of the ultrasound-velocity variation in the $C_{33}$ mode measured in pulsed magnetic fields. Since the anomalies at $B_m$ and $B^*$ are not observed in this mode, we also show a similar curve in the $C_T$ mode as a reference. The oscillations with frequency $A$ are clearly visible in $C_{33}$, both below and above $B^*$, even without subtracting a background. The presence of the frequency $A$ above $B^*$ is confirmed by performing the FFT on the background-subtracted data. The corresponding FFT spectra below and above $B^*$ are shown in Fig.~\ref{fig:MAQOC33pulsed}(b).

\begin{figure}[htb]
\includegraphics[width=\columnwidth]{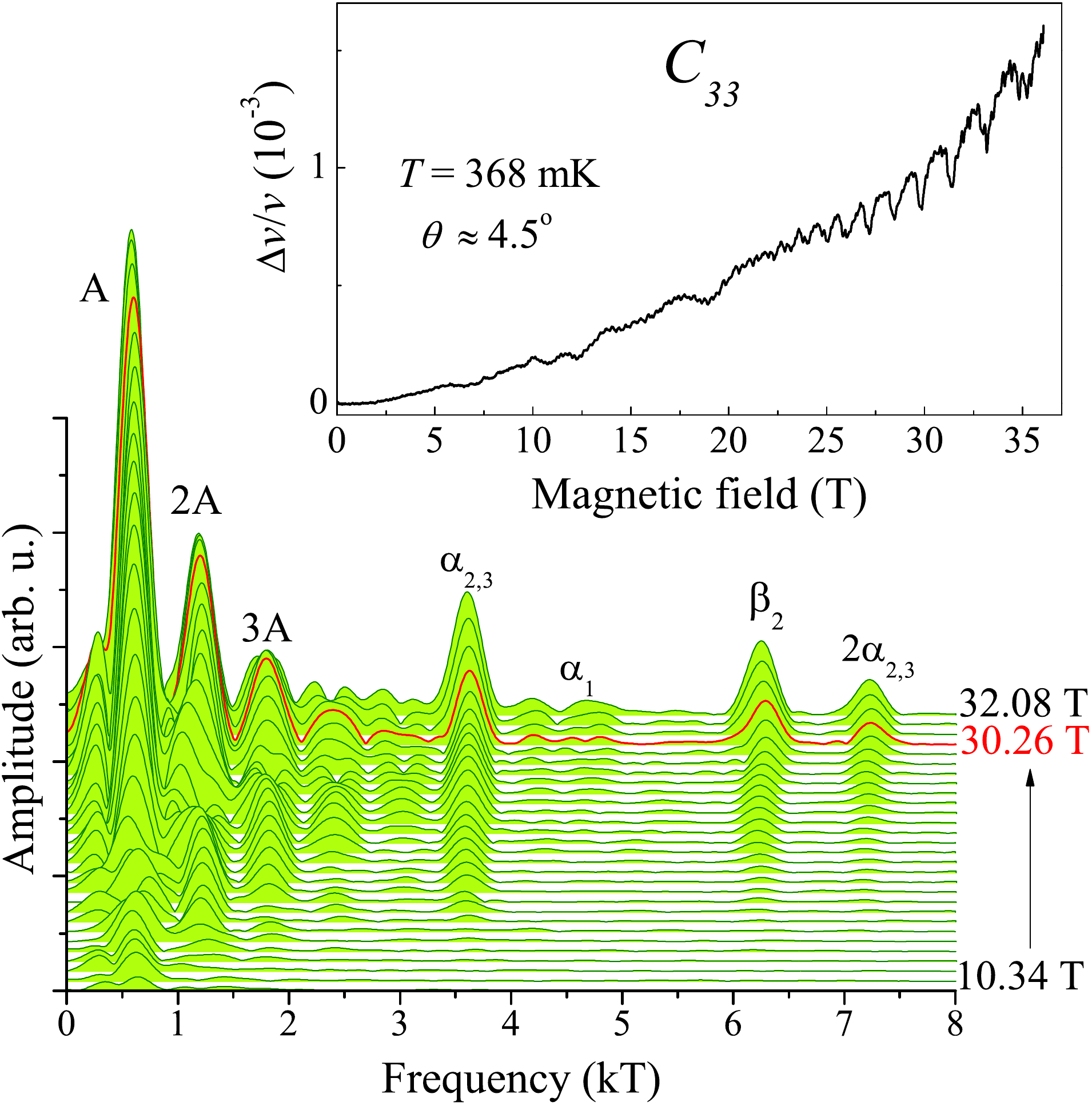}
\caption{\label{fig:MAQOC33static}FFT spectra of the static-field magnetoacoustic quantum oscillations, obtained by subtracting a nonoscillating background from the $\Delta v/v$ \emph{vs} $B$ curve in the $C_{33}$ mode shown in the inset, with $B$ tilted by 4.5$^\circ$ from the $c$ axis. The FFT spectra are obtained over the same $1/B$ range. For the bottom curve, the range is from $B_{min}$ = 10~T to $B_{max}$ = 10.71~T ($B_{avg}$ = 10.34~T). For each successive curve, $B_{min}$ is increased by 1~T up to 20~T, and by 0.5~T from there on. Curves are vertically shifted for clarity.}
\end{figure}

Our static-field data in the $C_{33}$ mode obtained at lower temperatures, an example of which is shown in the inset of Fig.~\ref{fig:MAQOC33static}, reveal many more quantum-oscillation frequencies. Some of the frequencies are clearly visible at as low field as 5~T. After subtracting a nonoscillating background from the data shown in the inset of Fig.~\ref{fig:MAQOC33static}, we performed FFTs over small equidistant $1/B$ ranges to keep the same frequency resolution. The resulting field dependence of the FFT spectra is shown in Fig.~\ref{fig:MAQOC33static}, where the field corresponding to the transition at $B^*$ is highlighted. It is clear from this FFT analysis that no frequencies emerge or disappear above $B^*$. This is in agrement with the results of our recent angle-dependent dHvA study~\cite{Mishra2021}. However, some of the dHvA frequencies observed at high fields in our previous study~\cite{Mishra2021} are not observed in our ultrasound-velocity data.

\section{Conclusions}

In summary, we performed high-field ultrasound-velocity measurements on bulk single crystals of CeRhIn$_5$. For a magnetic field slightly tilted away from the $c$ axis, we observed distinct ultrasound-velocity anomalies at both $B_m \simeq$~20~T and $B^* \simeq$~30~T at low temperatures in all symmetry-breaking modes, i.e., $C_{11}$, $C_{44}$, $C_{66}$, and $C_T$. In all these modes, the anomalies are of similar shape but of opposite sign at $B_m$ and $B^*$. Both anomalies are absent in the symmetry preserving $C_{33}$ mode. Furthermore, our temperature-dependent measurements reveal that both anomalies exist within the AFM state only. Given that the transition at $B_m$ corresponds to a change of magnetic structure from IC below $B_m$ to C above $B_m$, we argue that the transition at $B^*$ is of the same origin, i.e., from the C phase below $B^*$ to a new IC phase above it. This makes CeRhIn$_5$ one of the rare compounds, in which the application of a high magnetic field induces a C to IC transition. High-field neutron diffraction measurements would be of interest to definitely confirm our hypothesis.

\begin{acknowledgments}
We thank M.-B.~Lepetit, C.~Simon, R.~Settai, and Y.~Nemoto for fruitful discussions. We acknowledge the support of the LNCMI-CNRS, the HLD-HZDR, and the HFML-RU, members of the European Magnetic Field Laboratory (EMFL), the ANR-DFG grant ``Fermi-NESt'', the DFG through the W\"urzburg-Dresden Cluster of Excellence on Complexity and Topology in Quantum Matter -- $ct. qmat$  (EXC 2147, Project ID 39085490), and JSPS KAKENHI Grant Nos. JP15H05882, JP15H05884, JP15H05886, JP15K21732 (J-Physics).
\end{acknowledgments}

\bibliography{CeRhIn5_ultrasound}

\begin{thebibliography}{32}%
\makeatletter
\providecommand \@ifxundefined [1]{%
 \@ifx{#1\undefined}
}%
\providecommand \@ifnum [1]{%
 \ifnum #1\expandafter \@firstoftwo
 \else \expandafter \@secondoftwo
 \fi
}%
\providecommand \@ifx [1]{%
 \ifx #1\expandafter \@firstoftwo
 \else \expandafter \@secondoftwo
 \fi
}%
\providecommand \natexlab [1]{#1}%
\providecommand \enquote  [1]{``#1''}%
\providecommand \bibnamefont  [1]{#1}%
\providecommand \bibfnamefont [1]{#1}%
\providecommand \citenamefont [1]{#1}%
\providecommand \href@noop [0]{\@secondoftwo}%
\providecommand \href [0]{\begingroup \@sanitize@url \@href}%
\providecommand \@href[1]{\@@startlink{#1}\@@href}%
\providecommand \@@href[1]{\endgroup#1\@@endlink}%
\providecommand \@sanitize@url [0]{\catcode `\\12\catcode `\$12\catcode
  `\&12\catcode `\#12\catcode `\^12\catcode `\_12\catcode `\%12\relax}%
\providecommand \@@startlink[1]{}%
\providecommand \@@endlink[0]{}%
\providecommand \url  [0]{\begingroup\@sanitize@url \@url }%
\providecommand \@url [1]{\endgroup\@href {#1}{\urlprefix }}%
\providecommand \urlprefix  [0]{URL }%
\providecommand \Eprint [0]{\href }%
\providecommand \doibase [0]{https://doi.org/}%
\providecommand \selectlanguage [0]{\@gobble}%
\providecommand \bibinfo  [0]{\@secondoftwo}%
\providecommand \bibfield  [0]{\@secondoftwo}%
\providecommand \translation [1]{[#1]}%
\providecommand \BibitemOpen [0]{}%
\providecommand \bibitemStop [0]{}%
\providecommand \bibitemNoStop [0]{.\EOS\space}%
\providecommand \EOS [0]{\spacefactor3000\relax}%
\providecommand \BibitemShut  [1]{\csname bibitem#1\endcsname}%
\let\auto@bib@innerbib\@empty
%</preamble>
\bibitem [{\citenamefont {Gegenwart}\ \emph {et~al.}(2002)\citenamefont
  {Gegenwart}, \citenamefont {Custers}, \citenamefont {Geibel}, \citenamefont
  {Neumaier}, \citenamefont {Tayama}, \citenamefont {Tenya}, \citenamefont
  {Trovarelli},\ and\ \citenamefont {Steglich}}]{Gegenwart2002}%
  \BibitemOpen
  \bibfield  {author} {\bibinfo {author} {\bibfnamefont {P.}~\bibnamefont
  {Gegenwart}}, \bibinfo {author} {\bibfnamefont {J.}~\bibnamefont {Custers}},
  \bibinfo {author} {\bibfnamefont {C.}~\bibnamefont {Geibel}}, \bibinfo
  {author} {\bibfnamefont {K.}~\bibnamefont {Neumaier}}, \bibinfo {author}
  {\bibfnamefont {T.}~\bibnamefont {Tayama}}, \bibinfo {author} {\bibfnamefont
  {K.}~\bibnamefont {Tenya}}, \bibinfo {author} {\bibfnamefont
  {O.}~\bibnamefont {Trovarelli}},\ and\ \bibinfo {author} {\bibfnamefont
  {F.}~\bibnamefont {Steglich}},\ }\bibfield  {title} {\bibinfo {title}
  {{M}agnetic-{F}ield {I}nduced {Q}uantum {C}ritical {P}oint in
  $\mathrm{Y}\mathrm{b}\mathrm{R}{\mathrm{h}}_{\mathrm{2}}\mathrm{S}{\mathrm{i}}_{\mathrm{2}}$},\
  }\href {https://doi.org/10.1103/PhysRevLett.89.056402} {\bibfield  {journal}
  {\bibinfo  {journal} {Phys. Rev. Lett.}\ }\textbf {\bibinfo {volume} {89}},\
  \bibinfo {pages} {056402} (\bibinfo {year} {2002})}\BibitemShut {NoStop}%
\bibitem [{\citenamefont {Harrison}\ \emph {et~al.}(2007)\citenamefont
  {Harrison}, \citenamefont {Sebastian}, \citenamefont {Mielke}, \citenamefont
  {Paris}, \citenamefont {Gordon}, \citenamefont {Swenson}, \citenamefont
  {Rickel}, \citenamefont {Pacheco}, \citenamefont {Ruminer}, \citenamefont
  {Schillig}, \citenamefont {Sims}, \citenamefont {Lacerda}, \citenamefont
  {Suzuki}, \citenamefont {Harima},\ and\ \citenamefont
  {Ebihara}}]{Harrison2007}%
  \BibitemOpen
  \bibfield  {author} {\bibinfo {author} {\bibfnamefont {N.}~\bibnamefont
  {Harrison}}, \bibinfo {author} {\bibfnamefont {S.~E.}\ \bibnamefont
  {Sebastian}}, \bibinfo {author} {\bibfnamefont {C.~H.}\ \bibnamefont
  {Mielke}}, \bibinfo {author} {\bibfnamefont {A.}~\bibnamefont {Paris}},
  \bibinfo {author} {\bibfnamefont {M.~J.}\ \bibnamefont {Gordon}}, \bibinfo
  {author} {\bibfnamefont {C.~A.}\ \bibnamefont {Swenson}}, \bibinfo {author}
  {\bibfnamefont {D.~G.}\ \bibnamefont {Rickel}}, \bibinfo {author}
  {\bibfnamefont {M.~D.}\ \bibnamefont {Pacheco}}, \bibinfo {author}
  {\bibfnamefont {P.~F.}\ \bibnamefont {Ruminer}}, \bibinfo {author}
  {\bibfnamefont {J.~B.}\ \bibnamefont {Schillig}}, \bibinfo {author}
  {\bibfnamefont {J.~R.}\ \bibnamefont {Sims}}, \bibinfo {author}
  {\bibfnamefont {A.~H.}\ \bibnamefont {Lacerda}}, \bibinfo {author}
  {\bibfnamefont {M.-T.}\ \bibnamefont {Suzuki}}, \bibinfo {author}
  {\bibfnamefont {H.}~\bibnamefont {Harima}},\ and\ \bibinfo {author}
  {\bibfnamefont {T.}~\bibnamefont {Ebihara}},\ }\bibfield  {title} {\bibinfo
  {title} {Fermi {S}urface of {CeIn$_3$} above the {N\'{e}}el {C}ritical
  {F}ield},\ }\href {https://doi.org/10.1103/physrevlett.99.056401} {\bibfield
  {journal} {\bibinfo  {journal} {Phys. Rev. Lett.}\ }\textbf {\bibinfo
  {volume} {99}},\ \bibinfo {pages} {056401} (\bibinfo {year}
  {2007})}\BibitemShut {NoStop}%
\bibitem [{\citenamefont {L\'{e}vy}\ \emph {et~al.}(2005)\citenamefont
  {L\'{e}vy}, \citenamefont {Sheikin}, \citenamefont {Grenier},\ and\
  \citenamefont {Huxley}}]{Levy2005}%
  \BibitemOpen
  \bibfield  {author} {\bibinfo {author} {\bibfnamefont {F.}~\bibnamefont
  {L\'{e}vy}}, \bibinfo {author} {\bibfnamefont {I.}~\bibnamefont {Sheikin}},
  \bibinfo {author} {\bibfnamefont {B.}~\bibnamefont {Grenier}},\ and\ \bibinfo
  {author} {\bibfnamefont {A.~D.}\ \bibnamefont {Huxley}},\ }\bibfield  {title}
  {\bibinfo {title} {{Magnetic Field-Induced Superconductivity in the
  Ferromagnet URhGe}},\ }\href {https://doi.org/10.1126/science.1115498}
  {\bibfield  {journal} {\bibinfo  {journal} {Science}\ }\textbf {\bibinfo
  {volume} {309}},\ \bibinfo {pages} {1343} (\bibinfo {year}
  {2005})}\BibitemShut {NoStop}%
\bibitem [{\citenamefont {Aoki}\ \emph {et~al.}(2009)\citenamefont {Aoki},
  \citenamefont {D.~Matsuda}, \citenamefont {Taufour}, \citenamefont
  {Hassinger}, \citenamefont {Knebel},\ and\ \citenamefont
  {Flouquet}}]{Aoki2009}%
  \BibitemOpen
  \bibfield  {author} {\bibinfo {author} {\bibfnamefont {D.}~\bibnamefont
  {Aoki}}, \bibinfo {author} {\bibfnamefont {T.}~\bibnamefont {D.~Matsuda}},
  \bibinfo {author} {\bibfnamefont {V.}~\bibnamefont {Taufour}}, \bibinfo
  {author} {\bibfnamefont {E.}~\bibnamefont {Hassinger}}, \bibinfo {author}
  {\bibfnamefont {G.}~\bibnamefont {Knebel}},\ and\ \bibinfo {author}
  {\bibfnamefont {J.}~\bibnamefont {Flouquet}},\ }\bibfield  {title} {\bibinfo
  {title} {Extremely large and anisotropic upper critical field and the
  ferromagnetic instability in ucoge},\ }\href
  {https://doi.org/10.1143/JPSJ.78.113709} {\bibfield  {journal} {\bibinfo
  {journal} {J. Phys. Soc. Jpn.}\ }\textbf {\bibinfo {volume} {78}},\ \bibinfo
  {pages} {113709} (\bibinfo {year} {2009})}\BibitemShut {NoStop}%
\bibitem [{\citenamefont {Knebel}\ \emph {et~al.}(2019)\citenamefont {Knebel},
  \citenamefont {Knafo}, \citenamefont {Pourret}, \citenamefont {Niu},
  \citenamefont {Vališka}, \citenamefont {Braithwaite}, \citenamefont
  {Lapertot}, \citenamefont {Nardone}, \citenamefont {Zitouni}, \citenamefont
  {Mishra}, \citenamefont {Sheikin}, \citenamefont {Seyfarth}, \citenamefont
  {Brison}, \citenamefont {Aoki},\ and\ \citenamefont {Flouquet}}]{Knebel2019}%
  \BibitemOpen
  \bibfield  {author} {\bibinfo {author} {\bibfnamefont {G.}~\bibnamefont
  {Knebel}}, \bibinfo {author} {\bibfnamefont {W.}~\bibnamefont {Knafo}},
  \bibinfo {author} {\bibfnamefont {A.}~\bibnamefont {Pourret}}, \bibinfo
  {author} {\bibfnamefont {Q.}~\bibnamefont {Niu}}, \bibinfo {author}
  {\bibfnamefont {M.}~\bibnamefont {Vališka}}, \bibinfo {author}
  {\bibfnamefont {D.}~\bibnamefont {Braithwaite}}, \bibinfo {author}
  {\bibfnamefont {G.}~\bibnamefont {Lapertot}}, \bibinfo {author}
  {\bibfnamefont {M.}~\bibnamefont {Nardone}}, \bibinfo {author} {\bibfnamefont
  {A.}~\bibnamefont {Zitouni}}, \bibinfo {author} {\bibfnamefont
  {S.}~\bibnamefont {Mishra}}, \bibinfo {author} {\bibfnamefont
  {I.}~\bibnamefont {Sheikin}}, \bibinfo {author} {\bibfnamefont
  {G.}~\bibnamefont {Seyfarth}}, \bibinfo {author} {\bibfnamefont {J.-P.}\
  \bibnamefont {Brison}}, \bibinfo {author} {\bibfnamefont {D.}~\bibnamefont
  {Aoki}},\ and\ \bibinfo {author} {\bibfnamefont {J.}~\bibnamefont
  {Flouquet}},\ }\bibfield  {title} {\bibinfo {title} {{Field-Reentrant
  Superconductivity Close to a Metamagnetic Transition in the Heavy-Fermion
  Superconductor UTe$_2$}},\ }\href {https://doi.org/10.7566/JPSJ.88.063707}
  {\bibfield  {journal} {\bibinfo  {journal} {J. Phys. Soc. Jpn.}\ }\textbf
  {\bibinfo {volume} {88}},\ \bibinfo {pages} {063707} (\bibinfo {year}
  {2019})}\BibitemShut {NoStop}%
\bibitem [{\citenamefont {Daou}\ \emph {et~al.}(2006)\citenamefont {Daou},
  \citenamefont {Bergemann},\ and\ \citenamefont {Julian}}]{Daou2006}%
  \BibitemOpen
  \bibfield  {author} {\bibinfo {author} {\bibfnamefont {R.}~\bibnamefont
  {Daou}}, \bibinfo {author} {\bibfnamefont {C.}~\bibnamefont {Bergemann}},\
  and\ \bibinfo {author} {\bibfnamefont {S.~R.}\ \bibnamefont {Julian}},\
  }\bibfield  {title} {\bibinfo {title} {{Continuous Evolution of the Fermi
  Surface of ${\mathrm{CeRu}}_{2}{\mathrm{Si}}_{2}$ across the Metamagnetic
  Transition}},\ }\href {https://doi.org/10.1103/PhysRevLett.96.026401}
  {\bibfield  {journal} {\bibinfo  {journal} {Phys. Rev. Lett.}\ }\textbf
  {\bibinfo {volume} {96}},\ \bibinfo {pages} {026401} (\bibinfo {year}
  {2006})}\BibitemShut {NoStop}%
\bibitem [{\citenamefont {Rourke}\ \emph {et~al.}(2008)\citenamefont {Rourke},
  \citenamefont {McCollam}, \citenamefont {Lapertot}, \citenamefont {Knebel},
  \citenamefont {Flouquet},\ and\ \citenamefont {Julian}}]{Rourke2008}%
  \BibitemOpen
  \bibfield  {author} {\bibinfo {author} {\bibfnamefont {P.~M.~C.}\
  \bibnamefont {Rourke}}, \bibinfo {author} {\bibfnamefont {A.}~\bibnamefont
  {McCollam}}, \bibinfo {author} {\bibfnamefont {G.}~\bibnamefont {Lapertot}},
  \bibinfo {author} {\bibfnamefont {G.}~\bibnamefont {Knebel}}, \bibinfo
  {author} {\bibfnamefont {J.}~\bibnamefont {Flouquet}},\ and\ \bibinfo
  {author} {\bibfnamefont {S.~R.}\ \bibnamefont {Julian}},\ }\bibfield  {title}
  {\bibinfo {title} {{Magnetic-Field Dependence of the
  ${\mathrm{YbRh}}_{2}{\mathrm{Si}}_{2}$ Fermi Surface}},\ }\href
  {https://doi.org/10.1103/PhysRevLett.101.237205} {\bibfield  {journal}
  {\bibinfo  {journal} {Phys. Rev. Lett.}\ }\textbf {\bibinfo {volume} {101}},\
  \bibinfo {pages} {237205} (\bibinfo {year} {2008})}\BibitemShut {NoStop}%
\bibitem [{\citenamefont {Aoki}\ \emph {et~al.}(2016)\citenamefont {Aoki},
  \citenamefont {Seyfarth}, \citenamefont {Pourret}, \citenamefont {Gourgout},
  \citenamefont {McCollam}, \citenamefont {Bruin}, \citenamefont {Krupko},\
  and\ \citenamefont {Sheikin}}]{Aoki2016}%
  \BibitemOpen
  \bibfield  {author} {\bibinfo {author} {\bibfnamefont {D.}~\bibnamefont
  {Aoki}}, \bibinfo {author} {\bibfnamefont {G.}~\bibnamefont {Seyfarth}},
  \bibinfo {author} {\bibfnamefont {A.}~\bibnamefont {Pourret}}, \bibinfo
  {author} {\bibfnamefont {A.}~\bibnamefont {Gourgout}}, \bibinfo {author}
  {\bibfnamefont {A.}~\bibnamefont {McCollam}}, \bibinfo {author}
  {\bibfnamefont {J.~A.~N.}\ \bibnamefont {Bruin}}, \bibinfo {author}
  {\bibfnamefont {Y.}~\bibnamefont {Krupko}},\ and\ \bibinfo {author}
  {\bibfnamefont {I.}~\bibnamefont {Sheikin}},\ }\bibfield  {title} {\bibinfo
  {title} {Field-{I}nduced {L}ifshitz {T}ransition without {M}etamagnetism in
  {CeIrIn$_5$}},\ }\href {https://doi.org/10.1103/PhysRevLett.116.037202}
  {\bibfield  {journal} {\bibinfo  {journal} {Phys. Rev. Lett.}\ }\textbf
  {\bibinfo {volume} {116}},\ \bibinfo {pages} {037202} (\bibinfo {year}
  {2016})}\BibitemShut {NoStop}%
\bibitem [{\citenamefont {Hegger}\ \emph {et~al.}(2000)\citenamefont {Hegger},
  \citenamefont {Petrovic}, \citenamefont {Moshopoulou}, \citenamefont
  {Hundley}, \citenamefont {Sarrao}, \citenamefont {Fisk},\ and\ \citenamefont
  {Thompson}}]{Hegger2000}%
  \BibitemOpen
  \bibfield  {author} {\bibinfo {author} {\bibfnamefont {H.}~\bibnamefont
  {Hegger}}, \bibinfo {author} {\bibfnamefont {C.}~\bibnamefont {Petrovic}},
  \bibinfo {author} {\bibfnamefont {E.~G.}\ \bibnamefont {Moshopoulou}},
  \bibinfo {author} {\bibfnamefont {M.~F.}\ \bibnamefont {Hundley}}, \bibinfo
  {author} {\bibfnamefont {J.~L.}\ \bibnamefont {Sarrao}}, \bibinfo {author}
  {\bibfnamefont {Z.}~\bibnamefont {Fisk}},\ and\ \bibinfo {author}
  {\bibfnamefont {J.~D.}\ \bibnamefont {Thompson}},\ }\bibfield  {title}
  {\bibinfo {title} {Pressure-{I}nduced {S}uperconductivity in {Quasi-2D}
  {CeRhIn$_5$}},\ }\href {https://doi.org/10.1103/physrevlett.84.4986}
  {\bibfield  {journal} {\bibinfo  {journal} {Phys. Rev. Lett.}\ }\textbf
  {\bibinfo {volume} {84}},\ \bibinfo {pages} {4986} (\bibinfo {year}
  {2000})}\BibitemShut {NoStop}%
\bibitem [{\citenamefont {Bao}\ \emph {et~al.}(2000)\citenamefont {Bao},
  \citenamefont {Pagliuso}, \citenamefont {Sarrao}, \citenamefont {Thompson},
  \citenamefont {Fisk}, \citenamefont {Lynn},\ and\ \citenamefont
  {Erwin}}]{Bao2000}%
  \BibitemOpen
  \bibfield  {author} {\bibinfo {author} {\bibfnamefont {W.}~\bibnamefont
  {Bao}}, \bibinfo {author} {\bibfnamefont {P.~G.}\ \bibnamefont {Pagliuso}},
  \bibinfo {author} {\bibfnamefont {J.~L.}\ \bibnamefont {Sarrao}}, \bibinfo
  {author} {\bibfnamefont {J.~D.}\ \bibnamefont {Thompson}}, \bibinfo {author}
  {\bibfnamefont {Z.}~\bibnamefont {Fisk}}, \bibinfo {author} {\bibfnamefont
  {J.~W.}\ \bibnamefont {Lynn}},\ and\ \bibinfo {author} {\bibfnamefont
  {R.~W.}\ \bibnamefont {Erwin}},\ }\bibfield  {title} {\bibinfo {title}
  {Incommensurate magnetic structure of {CeRhIn$_5$}},\ }\href
  {https://doi.org/10.1103/physrevb.62.r14621} {\bibfield  {journal} {\bibinfo
  {journal} {Phys. Rev. B}\ }\textbf {\bibinfo {volume} {62}},\ \bibinfo
  {pages} {R14621} (\bibinfo {year} {2000})}\BibitemShut {NoStop}%
\bibitem [{\citenamefont {Fobes}\ \emph {et~al.}(2017)\citenamefont {Fobes},
  \citenamefont {Bauer}, \citenamefont {Thompson}, \citenamefont {Sazonov},
  \citenamefont {Hutanu}, \citenamefont {Zhang}, \citenamefont {Ronning},\ and\
  \citenamefont {Janoschek}}]{Fobes2017}%
  \BibitemOpen
  \bibfield  {author} {\bibinfo {author} {\bibfnamefont {D.~M.}\ \bibnamefont
  {Fobes}}, \bibinfo {author} {\bibfnamefont {E.~D.}\ \bibnamefont {Bauer}},
  \bibinfo {author} {\bibfnamefont {J.~D.}\ \bibnamefont {Thompson}}, \bibinfo
  {author} {\bibfnamefont {A.}~\bibnamefont {Sazonov}}, \bibinfo {author}
  {\bibfnamefont {V.}~\bibnamefont {Hutanu}}, \bibinfo {author} {\bibfnamefont
  {S.}~\bibnamefont {Zhang}}, \bibinfo {author} {\bibfnamefont
  {F.}~\bibnamefont {Ronning}},\ and\ \bibinfo {author} {\bibfnamefont
  {M.}~\bibnamefont {Janoschek}},\ }\bibfield  {title} {\bibinfo {title} {Low
  temperature magnetic structure of {CeRhIn$_5$} by neutron diffraction on
  absorption-optimized samples},\ }\href
  {https://doi.org/10.1088/1361-648x/aa6696} {\bibfield  {journal} {\bibinfo
  {journal} {J. Phys.: Condens. Matter}\ }\textbf {\bibinfo {volume} {29}},\
  \bibinfo {pages} {17LT01} (\bibinfo {year} {2017})}\BibitemShut {NoStop}%
\bibitem [{\citenamefont {Raymond}\ \emph {et~al.}(2007)\citenamefont
  {Raymond}, \citenamefont {Ressouche}, \citenamefont {Knebel}, \citenamefont
  {Aoki},\ and\ \citenamefont {Flouquet}}]{Raymond2007}%
  \BibitemOpen
  \bibfield  {author} {\bibinfo {author} {\bibfnamefont {S.}~\bibnamefont
  {Raymond}}, \bibinfo {author} {\bibfnamefont {E.}~\bibnamefont {Ressouche}},
  \bibinfo {author} {\bibfnamefont {G.}~\bibnamefont {Knebel}}, \bibinfo
  {author} {\bibfnamefont {D.}~\bibnamefont {Aoki}},\ and\ \bibinfo {author}
  {\bibfnamefont {J.}~\bibnamefont {Flouquet}},\ }\bibfield  {title} {\bibinfo
  {title} {Magnetic structure of {CeRhIn$_5$} under magnetic field},\ }\href
  {https://doi.org/10.1088/0953-8984/19/24/242204} {\bibfield  {journal}
  {\bibinfo  {journal} {J. Phys.: Condens. Matter}\ }\textbf {\bibinfo {volume}
  {19}},\ \bibinfo {pages} {242204} (\bibinfo {year} {2007})}\BibitemShut
  {NoStop}%
\bibitem [{\citenamefont {Fobes}\ \emph {et~al.}(2018)\citenamefont {Fobes},
  \citenamefont {Zhang}, \citenamefont {Lin}, \citenamefont {Das},
  \citenamefont {Ghimire}, \citenamefont {Bauer}, \citenamefont {Thompson},
  \citenamefont {Harriger}, \citenamefont {Ehlers}, \citenamefont {Podlesnyak},
  \citenamefont {Bewley}, \citenamefont {Sazonov}, \citenamefont {Hutanu},
  \citenamefont {Ronning}, \citenamefont {Batista},\ and\ \citenamefont
  {Janoschek}}]{Fobes2018}%
  \BibitemOpen
  \bibfield  {author} {\bibinfo {author} {\bibfnamefont {D.~M.}\ \bibnamefont
  {Fobes}}, \bibinfo {author} {\bibfnamefont {S.}~\bibnamefont {Zhang}},
  \bibinfo {author} {\bibfnamefont {S.-Z.}\ \bibnamefont {Lin}}, \bibinfo
  {author} {\bibfnamefont {P.}~\bibnamefont {Das}}, \bibinfo {author}
  {\bibfnamefont {N.~J.}\ \bibnamefont {Ghimire}}, \bibinfo {author}
  {\bibfnamefont {E.~D.}\ \bibnamefont {Bauer}}, \bibinfo {author}
  {\bibfnamefont {J.~D.}\ \bibnamefont {Thompson}}, \bibinfo {author}
  {\bibfnamefont {L.~W.}\ \bibnamefont {Harriger}}, \bibinfo {author}
  {\bibfnamefont {G.}~\bibnamefont {Ehlers}}, \bibinfo {author} {\bibfnamefont
  {A.}~\bibnamefont {Podlesnyak}}, \bibinfo {author} {\bibfnamefont {R.~I.}\
  \bibnamefont {Bewley}}, \bibinfo {author} {\bibfnamefont {A.}~\bibnamefont
  {Sazonov}}, \bibinfo {author} {\bibfnamefont {V.}~\bibnamefont {Hutanu}},
  \bibinfo {author} {\bibfnamefont {F.}~\bibnamefont {Ronning}}, \bibinfo
  {author} {\bibfnamefont {C.~D.}\ \bibnamefont {Batista}},\ and\ \bibinfo
  {author} {\bibfnamefont {M.}~\bibnamefont {Janoschek}},\ }\bibfield  {title}
  {\bibinfo {title} {Tunable emergent heterostructures in a prototypical
  correlated metal},\ }\href {https://doi.org/10.1038/s41567-018-0060-9}
  {\bibfield  {journal} {\bibinfo  {journal} {Nat. Phys.}\ }\textbf {\bibinfo
  {volume} {14}},\ \bibinfo {pages} {456} (\bibinfo {year} {2018})}\BibitemShut
  {NoStop}%
\bibitem [{\citenamefont {Mishra}\ \emph
  {et~al.}(2021{\natexlab{a}})\citenamefont {Mishra}, \citenamefont {Hornung},
  \citenamefont {Raba}, \citenamefont {Klotz}, \citenamefont {F\"{o}rster},
  \citenamefont {Harima}, \citenamefont {Aoki}, \citenamefont {Wosnitza},
  \citenamefont {McCollam},\ and\ \citenamefont {Sheikin}}]{Mishra2021}%
  \BibitemOpen
  \bibfield  {author} {\bibinfo {author} {\bibfnamefont {S.}~\bibnamefont
  {Mishra}}, \bibinfo {author} {\bibfnamefont {J.}~\bibnamefont {Hornung}},
  \bibinfo {author} {\bibfnamefont {M.}~\bibnamefont {Raba}}, \bibinfo {author}
  {\bibfnamefont {J.}~\bibnamefont {Klotz}}, \bibinfo {author} {\bibfnamefont
  {T.}~\bibnamefont {F\"{o}rster}}, \bibinfo {author} {\bibfnamefont
  {H.}~\bibnamefont {Harima}}, \bibinfo {author} {\bibfnamefont
  {D.}~\bibnamefont {Aoki}}, \bibinfo {author} {\bibfnamefont {J.}~\bibnamefont
  {Wosnitza}}, \bibinfo {author} {\bibfnamefont {A.}~\bibnamefont {McCollam}},\
  and\ \bibinfo {author} {\bibfnamefont {I.}~\bibnamefont {Sheikin}},\
  }\bibfield  {title} {\bibinfo {title} {{Robust Fermi-Surface Morphology of
  CeRhIn$_5$ across the Putative Field-Induced Quantum Critical Point}},\
  }\href {https://doi.org/10.1103/PhysRevLett.126.016403} {\bibfield  {journal}
  {\bibinfo  {journal} {Phys. Rev. Lett.}\ }\textbf {\bibinfo {volume} {126}},\
  \bibinfo {pages} {016403} (\bibinfo {year} {2021}{\natexlab{a}})}\BibitemShut
  {NoStop}%
\bibitem [{\citenamefont {Jiao}\ \emph {et~al.}(2015)\citenamefont {Jiao},
  \citenamefont {Chen}, \citenamefont {Kohama}, \citenamefont {Graf},
  \citenamefont {Bauer}, \citenamefont {Singleton}, \citenamefont {Zhu},
  \citenamefont {Weng}, \citenamefont {Pang}, \citenamefont {Shang},
  \citenamefont {Zhang}, \citenamefont {Lee}, \citenamefont {Park},
  \citenamefont {Jaime}, \citenamefont {Thompson}, \citenamefont {Steglich},
  \citenamefont {Si},\ and\ \citenamefont {Yuan}}]{Jiao2015}%
  \BibitemOpen
  \bibfield  {author} {\bibinfo {author} {\bibfnamefont {L.}~\bibnamefont
  {Jiao}}, \bibinfo {author} {\bibfnamefont {Y.}~\bibnamefont {Chen}}, \bibinfo
  {author} {\bibfnamefont {Y.}~\bibnamefont {Kohama}}, \bibinfo {author}
  {\bibfnamefont {D.}~\bibnamefont {Graf}}, \bibinfo {author} {\bibfnamefont
  {E.~D.}\ \bibnamefont {Bauer}}, \bibinfo {author} {\bibfnamefont
  {J.}~\bibnamefont {Singleton}}, \bibinfo {author} {\bibfnamefont {J.-X.}\
  \bibnamefont {Zhu}}, \bibinfo {author} {\bibfnamefont {Z.}~\bibnamefont
  {Weng}}, \bibinfo {author} {\bibfnamefont {G.}~\bibnamefont {Pang}}, \bibinfo
  {author} {\bibfnamefont {T.}~\bibnamefont {Shang}}, \bibinfo {author}
  {\bibfnamefont {J.}~\bibnamefont {Zhang}}, \bibinfo {author} {\bibfnamefont
  {H.-O.}\ \bibnamefont {Lee}}, \bibinfo {author} {\bibfnamefont
  {T.}~\bibnamefont {Park}}, \bibinfo {author} {\bibfnamefont {M.}~\bibnamefont
  {Jaime}}, \bibinfo {author} {\bibfnamefont {J.~D.}\ \bibnamefont {Thompson}},
  \bibinfo {author} {\bibfnamefont {F.}~\bibnamefont {Steglich}}, \bibinfo
  {author} {\bibfnamefont {Q.}~\bibnamefont {Si}},\ and\ \bibinfo {author}
  {\bibfnamefont {H.~Q.}\ \bibnamefont {Yuan}},\ }\bibfield  {title} {\bibinfo
  {title} {Fermi surface reconstruction and multiple quantum phase transitions
  in the antiferromagnet {CeRhIn$_5$}},\ }\href
  {https://doi.org/10.1073/pnas.1413932112} {\bibfield  {journal} {\bibinfo
  {journal} {Proc. Natl. Acad. Sci. USA}\ }\textbf {\bibinfo {volume} {112}},\
  \bibinfo {pages} {673} (\bibinfo {year} {2015})}\BibitemShut {NoStop}%
\bibitem [{\citenamefont {Jiao}\ \emph {et~al.}(2019)\citenamefont {Jiao},
  \citenamefont {Smidman}, \citenamefont {Kohama}, \citenamefont {Wang},
  \citenamefont {Graf}, \citenamefont {Weng}, \citenamefont {Zhang},
  \citenamefont {Matsuo}, \citenamefont {Bauer}, \citenamefont {Lee},
  \citenamefont {Kirchner}, \citenamefont {Singleton}, \citenamefont {Kindo},
  \citenamefont {Wosnitza}, \citenamefont {Steglich}, \citenamefont
  {Thompson},\ and\ \citenamefont {Yuan}}]{Jiao2019}%
  \BibitemOpen
  \bibfield  {author} {\bibinfo {author} {\bibfnamefont {L.}~\bibnamefont
  {Jiao}}, \bibinfo {author} {\bibfnamefont {M.}~\bibnamefont {Smidman}},
  \bibinfo {author} {\bibfnamefont {Y.}~\bibnamefont {Kohama}}, \bibinfo
  {author} {\bibfnamefont {Z.~S.}\ \bibnamefont {Wang}}, \bibinfo {author}
  {\bibfnamefont {D.}~\bibnamefont {Graf}}, \bibinfo {author} {\bibfnamefont
  {Z.~F.}\ \bibnamefont {Weng}}, \bibinfo {author} {\bibfnamefont {Y.~J.}\
  \bibnamefont {Zhang}}, \bibinfo {author} {\bibfnamefont {A.}~\bibnamefont
  {Matsuo}}, \bibinfo {author} {\bibfnamefont {E.~D.}\ \bibnamefont {Bauer}},
  \bibinfo {author} {\bibfnamefont {H.}~\bibnamefont {Lee}}, \bibinfo {author}
  {\bibfnamefont {S.}~\bibnamefont {Kirchner}}, \bibinfo {author}
  {\bibfnamefont {J.}~\bibnamefont {Singleton}}, \bibinfo {author}
  {\bibfnamefont {K.}~\bibnamefont {Kindo}}, \bibinfo {author} {\bibfnamefont
  {J.}~\bibnamefont {Wosnitza}}, \bibinfo {author} {\bibfnamefont
  {F.}~\bibnamefont {Steglich}}, \bibinfo {author} {\bibfnamefont {J.~D.}\
  \bibnamefont {Thompson}},\ and\ \bibinfo {author} {\bibfnamefont {H.~Q.}\
  \bibnamefont {Yuan}},\ }\bibfield  {title} {\bibinfo {title} {Enhancement of
  the effective mass at high magnetic fields in {CeRhIn$_5$}},\ }\href
  {https://doi.org/10.1103/physrevb.99.045127} {\bibfield  {journal} {\bibinfo
  {journal} {Phys. Rev. B}\ }\textbf {\bibinfo {volume} {99}},\ \bibinfo
  {pages} {045127} (\bibinfo {year} {2019})}\BibitemShut {NoStop}%
\bibitem [{\citenamefont {Moll}\ \emph {et~al.}(2015)\citenamefont {Moll},
  \citenamefont {Zeng}, \citenamefont {Balicas}, \citenamefont {Galeski},
  \citenamefont {Balakirev}, \citenamefont {Bauer},\ and\ \citenamefont
  {Ronning}}]{Moll2015}%
  \BibitemOpen
  \bibfield  {author} {\bibinfo {author} {\bibfnamefont {P.~J.~W.}\
  \bibnamefont {Moll}}, \bibinfo {author} {\bibfnamefont {B.}~\bibnamefont
  {Zeng}}, \bibinfo {author} {\bibfnamefont {L.}~\bibnamefont {Balicas}},
  \bibinfo {author} {\bibfnamefont {S.}~\bibnamefont {Galeski}}, \bibinfo
  {author} {\bibfnamefont {F.~F.}\ \bibnamefont {Balakirev}}, \bibinfo {author}
  {\bibfnamefont {E.~D.}\ \bibnamefont {Bauer}},\ and\ \bibinfo {author}
  {\bibfnamefont {F.}~\bibnamefont {Ronning}},\ }\bibfield  {title} {\bibinfo
  {title} {Field-induced density wave in the heavy-fermion compound
  {CeRhIn$_5$}},\ }\href {https://doi.org/10.1038/ncomms7663} {\bibfield
  {journal} {\bibinfo  {journal} {Nat. Commun.}\ }\textbf {\bibinfo {volume}
  {6}},\ \bibinfo {pages} {6663} (\bibinfo {year} {2015})}\BibitemShut
  {NoStop}%
\bibitem [{\citenamefont {Ronning}\ \emph {et~al.}(2017)\citenamefont
  {Ronning}, \citenamefont {Helm}, \citenamefont {Shirer}, \citenamefont
  {Bachmann}, \citenamefont {Balicas}, \citenamefont {Chan}, \citenamefont
  {Ramshaw}, \citenamefont {McDonald}, \citenamefont {Balakirev}, \citenamefont
  {Jaime}, \citenamefont {Bauer},\ and\ \citenamefont {Moll}}]{Ronning2017}%
  \BibitemOpen
  \bibfield  {author} {\bibinfo {author} {\bibfnamefont {F.}~\bibnamefont
  {Ronning}}, \bibinfo {author} {\bibfnamefont {T.}~\bibnamefont {Helm}},
  \bibinfo {author} {\bibfnamefont {K.~R.}\ \bibnamefont {Shirer}}, \bibinfo
  {author} {\bibfnamefont {M.~D.}\ \bibnamefont {Bachmann}}, \bibinfo {author}
  {\bibfnamefont {L.}~\bibnamefont {Balicas}}, \bibinfo {author} {\bibfnamefont
  {M.~K.}\ \bibnamefont {Chan}}, \bibinfo {author} {\bibfnamefont {B.~J.}\
  \bibnamefont {Ramshaw}}, \bibinfo {author} {\bibfnamefont {R.~D.}\
  \bibnamefont {McDonald}}, \bibinfo {author} {\bibfnamefont {F.~F.}\
  \bibnamefont {Balakirev}}, \bibinfo {author} {\bibfnamefont {M.}~\bibnamefont
  {Jaime}}, \bibinfo {author} {\bibfnamefont {E.~D.}\ \bibnamefont {Bauer}},\
  and\ \bibinfo {author} {\bibfnamefont {P.~J.~W.}\ \bibnamefont {Moll}},\
  }\bibfield  {title} {\bibinfo {title} {Electronic in-plane symmetry breaking
  at field-tuned quantum criticality in {CeRhIn$_5$}},\ }\href
  {https://doi.org/10.1038/nature23315} {\bibfield  {journal} {\bibinfo
  {journal} {Nature}\ }\textbf {\bibinfo {volume} {548}},\ \bibinfo {pages}
  {313} (\bibinfo {year} {2017})}\BibitemShut {NoStop}%
\bibitem [{\citenamefont {Takeuchi}\ \emph {et~al.}(2001)\citenamefont
  {Takeuchi}, \citenamefont {Inoue}, \citenamefont {Sugiyama}, \citenamefont
  {Aoki}, \citenamefont {Tokiwa}, \citenamefont {Haga}, \citenamefont {Kindo},\
  and\ \citenamefont {Ōnuki}}]{Takeuchi2001}%
  \BibitemOpen
  \bibfield  {author} {\bibinfo {author} {\bibfnamefont {T.}~\bibnamefont
  {Takeuchi}}, \bibinfo {author} {\bibfnamefont {T.}~\bibnamefont {Inoue}},
  \bibinfo {author} {\bibfnamefont {K.}~\bibnamefont {Sugiyama}}, \bibinfo
  {author} {\bibfnamefont {D.}~\bibnamefont {Aoki}}, \bibinfo {author}
  {\bibfnamefont {Y.}~\bibnamefont {Tokiwa}}, \bibinfo {author} {\bibfnamefont
  {Y.}~\bibnamefont {Haga}}, \bibinfo {author} {\bibfnamefont {K.}~\bibnamefont
  {Kindo}},\ and\ \bibinfo {author} {\bibfnamefont {Y.}~\bibnamefont
  {Ōnuki}},\ }\bibfield  {title} {\bibinfo {title} {{M}agnetic and {T}hermal
  {P}roperties of {CeIrIn$_5$} and {CeRhIn$_5$}},\ }\href
  {https://doi.org/10.1143/JPSJ.70.877} {\bibfield  {journal} {\bibinfo
  {journal} {J. Phys. Soc. Jpn.}\ }\textbf {\bibinfo {volume} {70}},\ \bibinfo
  {pages} {877} (\bibinfo {year} {2001})}\BibitemShut {NoStop}%
\bibitem [{\citenamefont {Rosa}\ \emph {et~al.}(2019)\citenamefont {Rosa},
  \citenamefont {Thomas}, \citenamefont {Balakirev}, \citenamefont {Bauer},
  \citenamefont {Fernandes}, \citenamefont {Thompson}, \citenamefont
  {Ronning},\ and\ \citenamefont {Jaime}}]{Rosa2019}%
  \BibitemOpen
  \bibfield  {author} {\bibinfo {author} {\bibfnamefont {P.}~\bibnamefont
  {Rosa}}, \bibinfo {author} {\bibfnamefont {S.}~\bibnamefont {Thomas}},
  \bibinfo {author} {\bibfnamefont {F.}~\bibnamefont {Balakirev}}, \bibinfo
  {author} {\bibfnamefont {E.}~\bibnamefont {Bauer}}, \bibinfo {author}
  {\bibfnamefont {R.}~\bibnamefont {Fernandes}}, \bibinfo {author}
  {\bibfnamefont {J.}~\bibnamefont {Thompson}}, \bibinfo {author}
  {\bibfnamefont {F.}~\bibnamefont {Ronning}},\ and\ \bibinfo {author}
  {\bibfnamefont {M.}~\bibnamefont {Jaime}},\ }\bibfield  {title} {\bibinfo
  {title} {Enhanced {H}ybridization {S}ets the {S}tage for {E}lectronic
  {N}ematicity in {CeRhIn$_5$}},\ }\href
  {https://doi.org/10.1103/physrevlett.122.016402} {\bibfield  {journal}
  {\bibinfo  {journal} {Phys. Rev. Lett.}\ }\textbf {\bibinfo {volume} {122}},\
  \bibinfo {pages} {016402} (\bibinfo {year} {2019})}\BibitemShut {NoStop}%
\bibitem [{\citenamefont {Lesseux}\ \emph {et~al.}(2020)\citenamefont
  {Lesseux}, \citenamefont {Sakai}, \citenamefont {Hattori}, \citenamefont
  {Tokunaga}, \citenamefont {Kambe}, \citenamefont {Kuhns}, \citenamefont
  {Reyes}, \citenamefont {Thompson}, \citenamefont {Pagliuso},\ and\
  \citenamefont {Urbano}}]{Lesseux2020}%
  \BibitemOpen
  \bibfield  {author} {\bibinfo {author} {\bibfnamefont {G.~G.}\ \bibnamefont
  {Lesseux}}, \bibinfo {author} {\bibfnamefont {H.}~\bibnamefont {Sakai}},
  \bibinfo {author} {\bibfnamefont {T.}~\bibnamefont {Hattori}}, \bibinfo
  {author} {\bibfnamefont {Y.}~\bibnamefont {Tokunaga}}, \bibinfo {author}
  {\bibfnamefont {S.}~\bibnamefont {Kambe}}, \bibinfo {author} {\bibfnamefont
  {P.~L.}\ \bibnamefont {Kuhns}}, \bibinfo {author} {\bibfnamefont {A.~P.}\
  \bibnamefont {Reyes}}, \bibinfo {author} {\bibfnamefont {J.~D.}\ \bibnamefont
  {Thompson}}, \bibinfo {author} {\bibfnamefont {P.~G.}\ \bibnamefont
  {Pagliuso}},\ and\ \bibinfo {author} {\bibfnamefont {R.~R.}\ \bibnamefont
  {Urbano}},\ }\bibfield  {title} {\bibinfo {title} {Orbitally defined
  field-induced electronic state in a {K}ondo lattice},\ }\href
  {https://doi.org/10.1103/PhysRevB.101.165111} {\bibfield  {journal} {\bibinfo
   {journal} {Phys. Rev. B}\ }\textbf {\bibinfo {volume} {101}},\ \bibinfo
  {pages} {165111} (\bibinfo {year} {2020})}\BibitemShut {NoStop}%
\bibitem [{\citenamefont {Jiao}\ \emph {et~al.}(2017)\citenamefont {Jiao},
  \citenamefont {Weng}, \citenamefont {Smidman}, \citenamefont {Graf},
  \citenamefont {Singleton}, \citenamefont {Bauer}, \citenamefont {Thompson},\
  and\ \citenamefont {Yuan}}]{Jiao2017}%
  \BibitemOpen
  \bibfield  {author} {\bibinfo {author} {\bibfnamefont {L.}~\bibnamefont
  {Jiao}}, \bibinfo {author} {\bibfnamefont {Z.~F.}\ \bibnamefont {Weng}},
  \bibinfo {author} {\bibfnamefont {M.}~\bibnamefont {Smidman}}, \bibinfo
  {author} {\bibfnamefont {D.}~\bibnamefont {Graf}}, \bibinfo {author}
  {\bibfnamefont {J.}~\bibnamefont {Singleton}}, \bibinfo {author}
  {\bibfnamefont {E.~D.}\ \bibnamefont {Bauer}}, \bibinfo {author}
  {\bibfnamefont {J.~D.}\ \bibnamefont {Thompson}},\ and\ \bibinfo {author}
  {\bibfnamefont {H.~Q.}\ \bibnamefont {Yuan}},\ }\bibfield  {title} {\bibinfo
  {title} {Magnetic field-induced {F}ermi surface reconstruction and quantum
  criticality in {CeRhIn$_5$}},\ }\href
  {https://doi.org/10.1080/14786435.2017.1282181} {\bibfield  {journal}
  {\bibinfo  {journal} {Philos. Mag.}\ }\textbf {\bibinfo {volume} {97}},\
  \bibinfo {pages} {3446} (\bibinfo {year} {2017})}\BibitemShut {NoStop}%
\bibitem [{\citenamefont {Mishra}\ \emph
  {et~al.}(2021{\natexlab{b}})\citenamefont {Mishra}, \citenamefont {Demuer},
  \citenamefont {Aoki},\ and\ \citenamefont {Sheikin}}]{Mishra2021a}%
  \BibitemOpen
  \bibfield  {author} {\bibinfo {author} {\bibfnamefont {S.}~\bibnamefont
  {Mishra}}, \bibinfo {author} {\bibfnamefont {A.}~\bibnamefont {Demuer}},
  \bibinfo {author} {\bibfnamefont {D.}~\bibnamefont {Aoki}},\ and\ \bibinfo
  {author} {\bibfnamefont {I.}~\bibnamefont {Sheikin}},\ }\bibfield  {title}
  {\bibinfo {title} {Specific heat of {CeRhIn$_5$} in high magnetic fields:
  {M}agnetic phase diagram revisited},\ }\href
  {https://doi.org/10.1103/PhysRevB.103.045110} {\bibfield  {journal} {\bibinfo
   {journal} {Phys. Rev. B}\ }\textbf {\bibinfo {volume} {103}},\ \bibinfo
  {pages} {045110} (\bibinfo {year} {2021}{\natexlab{b}})}\BibitemShut
  {NoStop}%
\bibitem [{\citenamefont {Kurihara}\ \emph {et~al.}(2020)\citenamefont
  {Kurihara}, \citenamefont {Miyake}, \citenamefont {Tokunaga}, \citenamefont
  {Hirose},\ and\ \citenamefont {Settai}}]{Kurihara2020}%
  \BibitemOpen
  \bibfield  {author} {\bibinfo {author} {\bibfnamefont {R.}~\bibnamefont
  {Kurihara}}, \bibinfo {author} {\bibfnamefont {A.}~\bibnamefont {Miyake}},
  \bibinfo {author} {\bibfnamefont {M.}~\bibnamefont {Tokunaga}}, \bibinfo
  {author} {\bibfnamefont {Y.}~\bibnamefont {Hirose}},\ and\ \bibinfo {author}
  {\bibfnamefont {R.}~\bibnamefont {Settai}},\ }\bibfield  {title} {\bibinfo
  {title} {High-field ultrasonic study of quadrupole ordering and crystal
  symmetry breaking in {CeRhIn$_5$}},\ }\href
  {https://doi.org/10.1103/PhysRevB.101.155125} {\bibfield  {journal} {\bibinfo
   {journal} {Phys. Rev. B}\ }\textbf {\bibinfo {volume} {101}},\ \bibinfo
  {pages} {155125} (\bibinfo {year} {2020})}\BibitemShut {NoStop}%
\bibitem [{\citenamefont {Shishido}\ \emph {et~al.}(2002)\citenamefont
  {Shishido}, \citenamefont {Settai}, \citenamefont {Aoki}, \citenamefont
  {Ikeda}, \citenamefont {Nakawaki}, \citenamefont {Nakamura}, \citenamefont
  {Iizuka}, \citenamefont {Inada}, \citenamefont {Sugiyama}, \citenamefont
  {Takeuchi}, \citenamefont {Kindo}, \citenamefont {Kobayashi}, \citenamefont
  {Haga}, \citenamefont {Harima}, \citenamefont {Aoki}, \citenamefont {Namiki},
  \citenamefont {Sato},\ and\ \citenamefont {{\={O}}nuki}}]{Shishido2002}%
  \BibitemOpen
  \bibfield  {author} {\bibinfo {author} {\bibfnamefont {H.}~\bibnamefont
  {Shishido}}, \bibinfo {author} {\bibfnamefont {R.}~\bibnamefont {Settai}},
  \bibinfo {author} {\bibfnamefont {D.}~\bibnamefont {Aoki}}, \bibinfo {author}
  {\bibfnamefont {S.}~\bibnamefont {Ikeda}}, \bibinfo {author} {\bibfnamefont
  {H.}~\bibnamefont {Nakawaki}}, \bibinfo {author} {\bibfnamefont
  {N.}~\bibnamefont {Nakamura}}, \bibinfo {author} {\bibfnamefont
  {T.}~\bibnamefont {Iizuka}}, \bibinfo {author} {\bibfnamefont
  {Y.}~\bibnamefont {Inada}}, \bibinfo {author} {\bibfnamefont
  {K.}~\bibnamefont {Sugiyama}}, \bibinfo {author} {\bibfnamefont
  {T.}~\bibnamefont {Takeuchi}}, \bibinfo {author} {\bibfnamefont
  {K.}~\bibnamefont {Kindo}}, \bibinfo {author} {\bibfnamefont {T.~C.}\
  \bibnamefont {Kobayashi}}, \bibinfo {author} {\bibfnamefont {Y.}~\bibnamefont
  {Haga}}, \bibinfo {author} {\bibfnamefont {H.}~\bibnamefont {Harima}},
  \bibinfo {author} {\bibfnamefont {Y.}~\bibnamefont {Aoki}}, \bibinfo {author}
  {\bibfnamefont {T.}~\bibnamefont {Namiki}}, \bibinfo {author} {\bibfnamefont
  {H.}~\bibnamefont {Sato}},\ and\ \bibinfo {author} {\bibfnamefont
  {Y.}~\bibnamefont {{\={O}}nuki}},\ }\bibfield  {title} {\bibinfo {title}
  {Fermi {S}urface, {M}agnetic and {S}uperconducting {P}roperties of
  {LaRhIn$_5$} and {CeTIn$_5$} ({T}: {C}o, {R}h and {I}r)},\ }\href
  {https://doi.org/10.1143/jpsj.71.162} {\bibfield  {journal} {\bibinfo
  {journal} {J. Phys. Soc. Jpn.}\ }\textbf {\bibinfo {volume} {71}},\ \bibinfo
  {pages} {162} (\bibinfo {year} {2002})}\BibitemShut {NoStop}%
\bibitem [{\citenamefont {L\"{u}thi}(2007)}]{Luethi2007}%
  \BibitemOpen
  \bibfield  {author} {\bibinfo {author} {\bibfnamefont {B.}~\bibnamefont
  {L\"{u}thi}},\ }\href {https://doi.org/10.1007/978-3-540-72194-9} {\emph
  {\bibinfo {title} {{Physical Acoustics in the Solid State}}}},\ Springer
  series in solid-state sciences\ (\bibinfo  {publisher} {Springer},\ \bibinfo
  {address} {Berlin, Heidelberg},\ \bibinfo {year} {2007})\BibitemShut
  {NoStop}%
\bibitem [{\citenamefont {Dan'shin}\ \emph {et~al.}(1987)\citenamefont
  {Dan'shin}, \citenamefont {Zherlitsyn}, \citenamefont {Zvada}, \citenamefont
  {Kramarchuk}, \citenamefont {Sdvizhkov},\ and\ \citenamefont
  {Fil'}}]{Danshin1987}%
  \BibitemOpen
  \bibfield  {author} {\bibinfo {author} {\bibfnamefont {N.~K.}\ \bibnamefont
  {Dan'shin}}, \bibinfo {author} {\bibfnamefont {S.~V.}\ \bibnamefont
  {Zherlitsyn}}, \bibinfo {author} {\bibfnamefont {S.~S.}\ \bibnamefont
  {Zvada}}, \bibinfo {author} {\bibfnamefont {G.~G.}\ \bibnamefont
  {Kramarchuk}}, \bibinfo {author} {\bibfnamefont {M.~A.}\ \bibnamefont
  {Sdvizhkov}},\ and\ \bibinfo {author} {\bibfnamefont {V.~D.}\ \bibnamefont
  {Fil'}},\ }\bibfield  {title} {\bibinfo {title} {Dynamic properties of
  {YbFeO$_3$} in the vicinity of an orientational phase transition},\
  }\href@noop {} {\bibfield  {journal} {\bibinfo  {journal} {Sov. Phys. JETP}\
  }\textbf {\bibinfo {volume} {66}},\ \bibinfo {pages} {1227} (\bibinfo {year}
  {1987})}\BibitemShut {NoStop}%
\bibitem [{\citenamefont {Fil}\ \emph {et~al.}(1991)\citenamefont {Fil},
  \citenamefont {Zzyagina}, \citenamefont {Zherlitsyn}, \citenamefont
  {Vitebsky}, \citenamefont {Sobolev}, \citenamefont {Barilo},\ and\
  \citenamefont {Zhigunov}}]{Fil1991}%
  \BibitemOpen
  \bibfield  {author} {\bibinfo {author} {\bibfnamefont {V.~D.}\ \bibnamefont
  {Fil}}, \bibinfo {author} {\bibfnamefont {G.~A.}\ \bibnamefont {Zzyagina}},
  \bibinfo {author} {\bibfnamefont {S.~V.}\ \bibnamefont {Zherlitsyn}},
  \bibinfo {author} {\bibfnamefont {I.~M.}\ \bibnamefont {Vitebsky}}, \bibinfo
  {author} {\bibfnamefont {V.~L.}\ \bibnamefont {Sobolev}}, \bibinfo {author}
  {\bibfnamefont {S.~N.}\ \bibnamefont {Barilo}},\ and\ \bibinfo {author}
  {\bibfnamefont {D.~I.}\ \bibnamefont {Zhigunov}},\ }\bibfield  {title}
  {\bibinfo {title} {Acoustiv properties of {Nd$_2$CuO$_4$} at low
  temperatures},\ }\href {https://doi.org/10.1142/S0217984991001672} {\bibfield
   {journal} {\bibinfo  {journal} {Modern Phys. Lett. B}\ }\textbf {\bibinfo
  {volume} {05}},\ \bibinfo {pages} {1367} (\bibinfo {year}
  {1991})}\BibitemShut {NoStop}%
\bibitem [{\citenamefont {Zherlitsyn}\ \emph {et~al.}(1993)\citenamefont
  {Zherlitsyn}, \citenamefont {Zvyagina}, \citenamefont {Fil'}, \citenamefont
  {Vitebskii}, \citenamefont {Barilo},\ and\ \citenamefont
  {Zhigunov}}]{Zherlitsyn1993}%
  \BibitemOpen
  \bibfield  {author} {\bibinfo {author} {\bibfnamefont {S.~V.}\ \bibnamefont
  {Zherlitsyn}}, \bibinfo {author} {\bibfnamefont {G.~A.}\ \bibnamefont
  {Zvyagina}}, \bibinfo {author} {\bibfnamefont {V.}~\bibnamefont {Fil'}},
  \bibinfo {author} {\bibfnamefont {I.~M.}\ \bibnamefont {Vitebskii}}, \bibinfo
  {author} {\bibfnamefont {S.~N.}\ \bibnamefont {Barilo}},\ and\ \bibinfo
  {author} {\bibfnamefont {D.}~\bibnamefont {Zhigunov}},\ }\bibfield  {title}
  {\bibinfo {title} {Magnetoelastic effects in {Nd$_{2-x}$Ce$_x$CuO$_4$} single
  crystals at low temperatures},\ }\href@noop {} {\bibfield  {journal}
  {\bibinfo  {journal} {Low. Temp. Phys.}\ }\textbf {\bibinfo {volume} {19}},\
  \bibinfo {pages} {934} (\bibinfo {year} {1993})}\BibitemShut {NoStop}%
\bibitem [{\citenamefont {Zherlitsyn}\ \emph {et~al.}(1995)\citenamefont
  {Zherlitsyn}, \citenamefont {Fil}, \citenamefont {Finsterbusch},
  \citenamefont {Molter}, \citenamefont {Wolf}, \citenamefont {Bruls},
  \citenamefont {Lüthi}, \citenamefont {Barilo},\ and\ \citenamefont
  {Zhigunov}}]{Zherlitsyn1995}%
  \BibitemOpen
  \bibfield  {author} {\bibinfo {author} {\bibfnamefont {S.}~\bibnamefont
  {Zherlitsyn}}, \bibinfo {author} {\bibfnamefont {V.}~\bibnamefont {Fil}},
  \bibinfo {author} {\bibfnamefont {D.}~\bibnamefont {Finsterbusch}}, \bibinfo
  {author} {\bibfnamefont {J.}~\bibnamefont {Molter}}, \bibinfo {author}
  {\bibfnamefont {B.}~\bibnamefont {Wolf}}, \bibinfo {author} {\bibfnamefont
  {G.}~\bibnamefont {Bruls}}, \bibinfo {author} {\bibfnamefont
  {B.}~\bibnamefont {Lüthi}}, \bibinfo {author} {\bibfnamefont
  {S.}~\bibnamefont {Barilo}},\ and\ \bibinfo {author} {\bibfnamefont
  {D.}~\bibnamefont {Zhigunov}},\ }\bibfield  {title} {\bibinfo {title} {High
  field phase diagram of {Nd$_{2-x}$Ce$_x$CuO$_4$} single crystals},\ }\href
  {https://doi.org/10.1016/0921-4526(94)00976-3} {\bibfield  {journal}
  {\bibinfo  {journal} {Phys. B: Condens. Matter}\ }\textbf {\bibinfo {volume}
  {211}},\ \bibinfo {pages} {168} (\bibinfo {year} {1995})}\BibitemShut
  {NoStop}%
\bibitem [{\citenamefont {Bachmann}\ \emph {et~al.}(2019)\citenamefont
  {Bachmann}, \citenamefont {Ferguson}, \citenamefont {Theuss}, \citenamefont
  {Meng}, \citenamefont {Putzke}, \citenamefont {Helm}, \citenamefont {Shirer},
  \citenamefont {Li}, \citenamefont {Modic}, \citenamefont {Nicklas},
  \citenamefont {K{\"o}nig}, \citenamefont {Low}, \citenamefont {Ghosh},
  \citenamefont {Mackenzie}, \citenamefont {Arnold}, \citenamefont {Hassinger},
  \citenamefont {McDonald}, \citenamefont {Winter}, \citenamefont {Bauer},
  \citenamefont {Ronning}, \citenamefont {Ramshaw}, \citenamefont {Nowack},\
  and\ \citenamefont {Moll}}]{Bachmann2019}%
  \BibitemOpen
  \bibfield  {author} {\bibinfo {author} {\bibfnamefont {M.~D.}\ \bibnamefont
  {Bachmann}}, \bibinfo {author} {\bibfnamefont {G.~M.}\ \bibnamefont
  {Ferguson}}, \bibinfo {author} {\bibfnamefont {F.}~\bibnamefont {Theuss}},
  \bibinfo {author} {\bibfnamefont {T.}~\bibnamefont {Meng}}, \bibinfo {author}
  {\bibfnamefont {C.}~\bibnamefont {Putzke}}, \bibinfo {author} {\bibfnamefont
  {T.}~\bibnamefont {Helm}}, \bibinfo {author} {\bibfnamefont {K.~R.}\
  \bibnamefont {Shirer}}, \bibinfo {author} {\bibfnamefont {Y.-S.}\
  \bibnamefont {Li}}, \bibinfo {author} {\bibfnamefont {K.~A.}\ \bibnamefont
  {Modic}}, \bibinfo {author} {\bibfnamefont {M.}~\bibnamefont {Nicklas}},
  \bibinfo {author} {\bibfnamefont {M.}~\bibnamefont {K{\"o}nig}}, \bibinfo
  {author} {\bibfnamefont {D.}~\bibnamefont {Low}}, \bibinfo {author}
  {\bibfnamefont {S.}~\bibnamefont {Ghosh}}, \bibinfo {author} {\bibfnamefont
  {A.~P.}\ \bibnamefont {Mackenzie}}, \bibinfo {author} {\bibfnamefont
  {F.}~\bibnamefont {Arnold}}, \bibinfo {author} {\bibfnamefont
  {E.}~\bibnamefont {Hassinger}}, \bibinfo {author} {\bibfnamefont {R.~D.}\
  \bibnamefont {McDonald}}, \bibinfo {author} {\bibfnamefont {L.~E.}\
  \bibnamefont {Winter}}, \bibinfo {author} {\bibfnamefont {E.~D.}\
  \bibnamefont {Bauer}}, \bibinfo {author} {\bibfnamefont {F.}~\bibnamefont
  {Ronning}}, \bibinfo {author} {\bibfnamefont {B.~J.}\ \bibnamefont
  {Ramshaw}}, \bibinfo {author} {\bibfnamefont {K.~C.}\ \bibnamefont
  {Nowack}},\ and\ \bibinfo {author} {\bibfnamefont {P.~J.~W.}\ \bibnamefont
  {Moll}},\ }\bibfield  {title} {\bibinfo {title} {Spatial control of
  heavy-fermion superconductivity in {CeIrIn$_5$}},\ }\href
  {https://doi.org/10.1126/science.aao6640} {\bibfield  {journal} {\bibinfo
  {journal} {Science}\ }\textbf {\bibinfo {volume} {366}},\ \bibinfo {pages}
  {221} (\bibinfo {year} {2019})}\BibitemShut {NoStop}%
\bibitem [{\citenamefont {Yashima}\ \emph {et~al.}(2020)\citenamefont
  {Yashima}, \citenamefont {Michizoe}, \citenamefont {Mukuda}, \citenamefont
  {Shishido}, \citenamefont {Settai},\ and\ \citenamefont
  {{\={O}}nuki}}]{Yashima2020}%
  \BibitemOpen
  \bibfield  {author} {\bibinfo {author} {\bibfnamefont {M.}~\bibnamefont
  {Yashima}}, \bibinfo {author} {\bibfnamefont {R.}~\bibnamefont {Michizoe}},
  \bibinfo {author} {\bibfnamefont {H.}~\bibnamefont {Mukuda}}, \bibinfo
  {author} {\bibfnamefont {H.}~\bibnamefont {Shishido}}, \bibinfo {author}
  {\bibfnamefont {R.}~\bibnamefont {Settai}},\ and\ \bibinfo {author}
  {\bibfnamefont {Y.}~\bibnamefont {{\={O}}nuki}},\ }\bibfield  {title}
  {\bibinfo {title} {{Incommensurate Antiferromagnetic Order under Pressure in
  CeRhIn$_5$ Studied by $^{115}$In-NQR}},\ }\href
  {https://doi.org/10.7566/jpscp.29.011010} {\bibfield  {journal} {\bibinfo
  {journal} {JPS Conf. Proc.}\ }\textbf {\bibinfo {volume} {29}},\ \bibinfo
  {pages} {011010} (\bibinfo {year} {2020})}\BibitemShut {NoStop}%
\end{thebibliography}%

\end{document}